\documentstyle[11pt,epsf]{article}
\newtheorem{theorem}{Theorem}

\newcommand {\dfn} {\stackrel{\Delta} {=}}
\newcommand {\exe} {\stackrel{\cdot} {=}}
\newcommand {\gexe} {\stackrel{\cdot} {\ge}}
\newcommand {\lexe} {\stackrel{\cdot} {\le}}
\newcommand{\eqa}{\stackrel{\mbox{(a)}}{=}}

\newcommand{\leb}{\stackrel{\mbox{(b)}}{\le}}

\newcommand {\bi} {\mbox{\boldmath $i$}}

\newcommand {\bx} {\mbox{\boldmath $x$}}
\newcommand {\by} {\mbox{\boldmath $y$}}
\newcommand {\bz} {\mbox{\boldmath $z$}}

\newcommand {\bE} {\mbox{\boldmath $E$}}

\newcommand {\hP} {\hat{P}}
\newcommand {\hI} {\hat{I}}
\newcommand {\hH} {\hat{H}}
\newcommand {\tQ} {\tilde{Q}}

\newcommand {\tV} {\tilde{V}}

\newcommand {\bX} {\mbox{\boldmath $X$}}
\newcommand {\bY} {\mbox{\boldmath $Y$}}

\newcommand{\calC}{{\cal C}}

\newcommand{\calE}{{\cal E}}

\newcommand{\calG}{{\cal G}}

\newcommand{\calI}{{\cal I}}

\newcommand{\calM}{{\cal M}}

\newcommand{\calQ}{{\cal Q}}

\newcommand{\calS}{{\cal S}}
\newcommand{\calT}{{\cal T}}
\newcommand{\calU}{{\cal U}}

\newcommand{\calX}{{\cal X}}
\newcommand{\calY}{{\cal Y}}
\newcommand{\calZ}{{\cal Z}}

\begin{document}
\thispagestyle{empty}
\title{Reliability of Universal Decoding\\
Based on Vector--Quantized Codewords}
\author{Neri Merhav}
\date{}
\maketitle

\begin{center}
The Andrew \& Erna Viterbi Faculty of Electrical Engineering\\
Technion - Israel Institute of Technology \\
Technion City, Haifa 32000, ISRAEL \\
E--mail: {\tt merhav@ee.technion.ac.il}\\
\end{center}
\vspace{1.5\baselineskip}
\setlength{\baselineskip}{1.5\baselineskip}

\begin{abstract}
Motivated by applications of biometric identification and content
identification systems,
we consider the problem of random
coding for channels, where each codeword undergoes lossy compression (vector
quantization), and where
the decoder bases its decision only on the compressed codewords and the
channel output, which is in turn, the channel's response to the transmission of
an original codeword, before
compression. For memoryless
sources and memoryless channels with finite
alphabets, we propose a new universal decoder and analyze its error
exponent, which improves on an earlier result by Dasarathy and Draper (2011),
who used the classic maximum mutual information (MMI) universal decoder.
Further, we show that our universal decoder provides the same error exponent
as that of the optimal, maximum likelihood (ML) decoder, at least as long as
all single--letter transition probabilities of the channel are positive.
We conjecture that the same argument remains true even without this positivity
condition.\\

\end{abstract}

\newpage
\section{Introduction}

The problems of biometric identification (see, e.g., \cite[Chap.\ 5]{IW10},
\cite{Tuncel09}, \cite{WKGL03}, \cite{WKBL} and references therein)
and content identification (\cite{DD11}, \cite{DD14} see also the related
problem of pattern recognition \cite{WO08})
have received some considerable attention in the last few years.

Both of these problems have a certain version that, in a nutshell, can be
described in two phases, as follows. In the first phase, a.k.a.\ the {\it
enrollment phase}, $e^{nR}$ mutually independent, randomly drawn vectors of length $n$ are
quantized and stored in a database. In the second phases, a.k.a.\ the {\it
identification phase}, a noisy version of one of the original random
vectors (before quantization) is presented to the system, which in turn has to
identify the index of the corresponding stored (compressed)
vector. In the case of biometric identification systems, the various signals are
biometric ones (e.g., voices, fingerprints, face photographs, irises, etc.) 
corresponding to a group of individuals who subscribe to the
biometric system in the enrollment phase, 
and the storage of these signals (which are naturally analog in their original
form), using a finite amount of memory, can be
carried out, of course, within finite accuracy only, due to the quantization. 
In content identification, the
scenario is similar except that the various signals represent contents (for
example, documents, images or video files \cite{VW09}), 
which are desired to be identified (in spite of some possible 
modifications) and found in the system,
whenever existent therein.

From the information--theoretic point of view, 
this problem naturally falls within the framework of coded communication in the
random coding regime,\footnote{While in classic information theory, the
concept of random coding is, first and foremost, a trick for a
non--constructive proof for the existence of good codes, here it is part of
the model, which represents the biometric source, or the source that generates
the contents, depending on the application.}
where the decoder does not have direct access to the
original transmitted codewords themselves, but only to distorted versions of these codewords,
that are obtained after lossy compression. Nonetheless, the channel output that is
presented to the decoder is obtained as the response of the channel to
one of the original codewords, before the lossy compression.
For a memoryless source and channel,
the maximum achievable rate $R$ (i.e., the capacity) of this model setting has already
been established by Tuncel \cite{Tuncel09} (see also \cite{WO08},
\cite{WKGL03}, \cite{WKBL}). Two years later, 
Dasarathy and Draper \cite{DD11} have derived a lower bound to the achievable
reliability (achievable error exponent) at a given rate $R$, and then after three
more years \cite{DD14},
the same authors have also derived an upper (converse) bound to
the reliability function based on a sphere--packing argument.

In this paper, we improve on the analysis in \cite{DD11}. In
particular, while Dasarathy and Draper chose to analyze the 
performance of the well--known maximum
mutual information (MMI) decoder \cite{CK11}, without an apparent explanation and
justification for this choice of decoder, 
here we argue that, in this special setting, there is
room for improvement over the MMI decoder, in two different aspects. The first is
relevant even without lossy compression: the MMI decoding metric is universally
optimal (in the sense of the random coding error exponent) when the code ensemble
is defined by the uniform distribution within a given type class, but
when the random coding distribution is
i.i.d.\ (as in the model considered in \cite{DD11} and here), the MMI decoding
metric should be modified by adding a divergence term between the empirical
distribution of the codeword being tested and the true random coding distribution
(see \cite[eq.\ (16)]{givenclassofmetrics}). On top of that, when the compression ingredient
is brought back into the picture, this divergence term should be modified too.
The second aspect of the improvement over the MMI decoder, is that the MMI
metric should also be modified to account to the fact
that after compression, the support of the induced random coding 
distribution is limited to the reproduction codebook of the lossy source
encoder. As a consequence, instead of the normalized log--cardinality of the conditional type of
the codeword given the channel output (which appears in the analysis of the usual setting and 
yields the conditional empirical
entropy term that is part of the MMI metric), it turns out that one should
better use
the normalized logarithm of the {\it number of reproduction vectors that are
jointly typical} with the channel output. 

The main part of this paper is in the performance analysis of a new universal
decoder that is obtained after the two above described modifications, and our
main contributions are as follows.
\begin{enumerate}
\item Exponentially tight error performance analysis for the new proposed universal decoder.
\item Comparison with the result in \cite{DD11}. The error exponent of the
proposed decoder is at least as large as that of \cite{DD11}, and often,
strictly so.
\item It is shown that the new universal decoder provides the same random coding error exponent
as the optimal maximum likelihood (ML) decoder at least as long as all
single--letter transition probabilities of the channel are positive. We
believe that this positivity limitation is merely a technical issue, and in
fact, this finding continues to hold true even without this limitation.
The source of this belief is the fact that random coding exponents are
normally continuous in the channel parameters.
\item The new proposed decoder is shown to be no worse than any other decoder
that bases its decision solely on the joint empirical distribution of the
codebook vector being tested and the channel output, and this holds for any
memoryless channel, even without the positivity limitation mentioned in item 3.
\item As a byproduct of the above, we also provide a good approximation to the
ML decoder that is based on empirical distributions only (in the sense of item
4). This approximation applies to the vast majority of lossy compression
codebooks in the ensemble, as long as the channel satisfies the positivity condition.
The approximation could be useful because even when the channel is known, 
the exact ML decoder is hard to
implement, due to the compression part.
\end{enumerate}

The outline of the remaining part of this paper is as follows.
In Section 2, we establish notation conventions. Section 3 is devoted to the formal
description of the problem. Section 4 provides an informal outline of the
basic idea of
this work. In Section 5, we formally introduce the proposed universal decoder,
and then, state and prove the main result of this work, along with a
discussion that contains, among other things, a comparison
with \cite{DD11}.
In Section 6, we derive a matching lower bound to the average error
probability of the ML decoder.
Finally, in Section 7, we summarize and conclude.

\section{Notation Conventions}

Throughout the paper, random variables will be denoted by capital
letters, specific values they may take will be denoted by the
corresponding lower case letters, and their alphabets
will be denoted by calligraphic letters. Random
vectors and their realizations will be denoted,
respectively, by capital letters and the corresponding lower case letters,
both in the bold face font. Their alphabets will be superscripted by their
dimensions. For example, the random vector $\bX=(X_1,\ldots,X_n)$, ($n$ --
positive integer) may take a specific vector value $\bx=(x_1,\ldots,x_n)$
in $\calX^n$, the $n$--th order Cartesian power of $\calX$, which is
the alphabet of each component of this vector.
Sources and channels will be 
subscripted by the names of the relevant random variables/vectors and their
conditionings, whenever needed and if applicable, following the standard notation conventions,
e.g., $Q_X$, $Q_{Y|X}$, and so on. When there is no room for ambiguity, these
subscripts will be omitted. For a given $Q_X$ and $Q_{Y|X}$, the notation
$(Q_X\times Q_{Y|X})_Y$ will be used to denote the operation that returns the
induced marginal of $Y$, that is, $Q_Y(y)=\sum_{x\in\calX}Q_X(x)Q_{Y|X}(y|x)$,
and a similar notation rule will apply to other pairs (or triples) of random
variables. For a generic joint distribution
$Q_{XY}=\{Q_{XY}(x,y),~x\in\calX,~y\in\calY\}$, which will often be abbreviated
by $Q$, information measures will be denoted in the conventional manner, but
with a subscript $Q$, that is $H_Q(X)$ is the marginal entropy of $X$,
$H_Q(X|Y)$ is the conditional entropy of $X$ given $Y$,
$I_Q(X;Y)=H_Q(X)-H_Q(X|Y)$ is the mutual mutual information, $D(Q_X\|G)$ is
the relative entropy between $Q_X$ and another distribution
$G=\{G(x),~x\in\calX\}$, and so on. The weighted divergence between two
conditional distributions (channels), say, $Q_{Z|X}$ and
$W=\{W(z|x),x\in\calX,~z\in\calZ\}$, with weighting $Q_X$ is defined as
\begin{equation}
D(Q_{Z|X}\|W|Q_X)=\sum_{x\in\calX}Q_X(x)\sum_{z\in\calZ}Q_{Z|X}(z|x)\log\frac{Q_{Z|X}(z|x)}{W(z|x)}.
\end{equation}
The probability of an event $\calE$ under $P$ will be denoted by $P[\calE]$,
and the expectation
operator with respect to (w.r.t.) a probability distribution $P$ will be
denoted by
$\bE_P\{\cdot\}$. Again, the subscript will be omitted if the underlying
probability distribution is clear from the context.
For two positive sequences $a_n$ and $b_n$, the notation $a_n\exe b_n$ will
stand for equality in the exponential scale, that is,
$\lim_{n\to\infty}\frac{1}{n}\log \frac{a_n}{b_n}=0$. Similarly,
$a_n\lexe b_n$ means that
$\limsup_{n\to\infty}\frac{1}{n}\log \frac{a_n}{b_n}\le 0$, and so on.
The indicator function
of an event $\calE$ will be denoted by $\calI\{E\}$. The notation $[x]_+$
will stand for $\max\{0,x\}$.

The empirical distribution of a sequence $\bx\in\calX^n$, which will be
denoted by $\hat{P}_{\bx}$, is the vector of relative frequencies
$\hat{P}_{\bx}(x)$
of each symbol $x\in\calX$ in $\bx$.
The type class of $\bx\in\calX^n$, denoted $\calT(\bx)$, is the set of all
vectors $\bx'$
with $\hat{P}_{\bx'}=\hat{P}_{\bx}$. When we wish to emphasize the
dependence of the type class on the empirical distribution $\hat{P}$, we
will denote it by
$\calT(\hat{P})$. Information measures associated with empirical distributions
will be denoted with `hats' and will be subscripted by the sequences from
which they are induced. For example, the entropy associated with
$\hat{P}_{\bx}$, which is the empirical entropy of $\bx$, will be denoted by
$\hat{H}_{\bx}(X)$. 
Similar conventions will apply to the joint empirical
distribution, the joint type class, the conditional empirical distributions
and the conditional type classes associated with pairs (and multiples) of
sequences of length $n$.
Accordingly, $\hP_{\bx\by}$ would be the joint empirical
distribution of $(\bx,\by)=\{(x_i,y_i)\}_{i=1}^n$,
$\calT(\bx,\by)$ or $\calT(\hP_{\bx\by})$ will denote
the joint type class of $(\bx,\by)$, $\calT(\bx|\by)$ will stand for
the conditional type class of $\bx$ given
$\by$, $\hH_{\bx\by}(X,Y)$ will designate the empirical joint entropy of $\bx$
and $\by$,
$\hH_{\bx\by}(X|Y)$ will be the empirical conditional entropy,
$\hI_{\bx\by}(X;Y)$ will
denote empirical mutual information, and so on. When we wish to emphasize the
dependence of $\calT(\bx|\by)$ upon $\by$ and the relevant empirical conditional
distribution, $Q_{X|Y}=\hP_{\bx|\by}$, we denote it by $\calT(Q_{X|Y}|\by)$.
Similar conventions will apply to triples of sequences, say, $\{(\bx,\by,\bz)\}$,
etc. Likewise, when we wish to emphasize the dependence of empirical information
measures upon a given empirical distribution given by $Q$, we denote them
using the subscript $Q$, as described above.

\section{Problem Formulation}

\subsection{General Setting}

Consider a discrete memoryless source (DMS), $G$, which, in the enrollment
phase, generates
$M=e^{nR_{\mbox{\tiny I}}}$ vectors of length $n$, $\bx_1,\ldots,\bx_M$, $\bx_m\in\calX^n$,
$m=1,2,\ldots,M$, $\calX^n$ being the $n$--th Cartesian power of a finite
alphabet $\calX$, and $R_{\mbox{\tiny I}}$ being the identification rate.
Each such vector is generated according to
\begin{equation}
G(\bx)=\prod_{i=1}^nG(x_i),
\end{equation}
where $G=\{G(x),~x\in\calX\}$ designates the source.
To complete the enrollment phase, each vector $\bx_m$, $m=1,2,\ldots,M$, 
is fed into a lossy source encoder (vector quantizer), whose output is 
$\by_m=f(\bx_m)\in\calY^n$ (the $n$--th Cartesian power of another
finite alphabet, $\calY$), and then $\by_m$ is stored
in the database. The construction of $f(\cdot)$,
which must trade off between compression constraints and identification
performance,
will be described in Subsection 3.2.

In the identification phase, an index $m$ is selected uniformly at random
and then a noisy version $\bz$, of $\bx_m$, is presented to the system with
the query to identify $m$, based on $\bz$ and on the codebook 
$\calC=\{\by_1,\ldots,\by_M\}$ of quantized enrollment vectors. 
This noisy version $\bz\in\calZ^n$ ($\calZ^n$ being the $n$--th Cartesian
power of yet another finite alphabet, $\calZ$),
is generated by a discrete memoryless channel (DMC),
according to $W(\bz|\bx_m)$, where for a generic $\bx\in\calX^n$,
\begin{equation}
\label{wchannel}
W(\bz|\bx)=\prod_{i=1}^n W(z_i|x_i),
\end{equation}
and we denote by $W$ the matrix of the single--letter transition 
probabilities, $\{W(z|x),~x\in\calX,~z\in\calZ\}$.

As in \cite{DD11}, we are interested in an achievable exponential bound
to the error probability in decoding the index $m$ for the query in the
identification phase. In principle, the problem 
falls in the ordinary framework of ML decoding with the likelihood function
\begin{equation}
\label{exactlikelihood}
P(\bz|\by_m)=\frac{P(\by_m,\bz)}{P(\by_m)}=
\frac{\sum_{\bx\in\calX^n}G(\bx)W(\bz|\bx)\calI\{\bx\in f^{-1}(\by_m)\}}
{\sum_{\bx\in\calX^n}G(\bx)\calI\{\bx\in f^{-1}(\by_m)\}},
\end{equation}
where $f^{-1}(\by_m)=\{\bx\in\calX^n:~f(\bx)=\by_m\}$ is the inverse image
of $\by_m$ induced by the lossy encoder $f$.
We would like to characterize an ensemble of
source encoders $\{f\}$, that satisfy a certain 
compression constraint, and a universal decoder $\hat{m}=g(\bz,\calC)$,
whose average (over the ensemble of $\{f\}$) error probability,
\begin{equation}
\bar{\mbox{P}}_{\mbox{\tiny e}}=\frac{1}{M}\sum_{m=1}^M
\mbox{Pr}\{g(\bz,\calC)\ne m\},
\end{equation}
is as small as possible, or more precisely, its
error exponent,
\begin{equation}
E(R)=\lim_{n\to\infty}\left[-\frac{\log \bar{\mbox{P}}_{\mbox{\tiny
e}}}{n}\right],
\end{equation}
is as large as possible (provided that the limit exists).

Let $L$ be a length function of a lossless code, that is, a function from
$\calC$ to the positive integers, satisfying the Kraft inequality,
$\sum_{\by\in\calC}2^{-L(\by)}\le 1$. Also, let $R_{\mbox{\tiny C}} > 0$ be given.
The compression constraint can be formalized 
in many ways. A few examples are the following.
\begin{enumerate}
\item {\it Expected length constraint:} $\bE\{L(\bY)\}\le nR_{\mbox{\tiny C}}$.
\item {\it Excess--length probability constraint:} $\mbox{Pr}\{L(\bY)\ge
nR_{\mbox{\tiny C}}\}\le e^{-nE_{\mbox{\tiny C}}}$ for a given $E_{\mbox{\tiny
C}} > 0$.
\item {\it Exponential moment constraint:} $\bE\{\exp[sL(\bY)]\}\le
e^{n\Lambda}$ for given $s>0$ and $\Lambda > 0$.
\end{enumerate}

\subsection{The Ensemble of Lossy Encoders}

We now move on to describe the construction of lossy encoder
$f:\calX^n\to\calC$, or more
precisely, the ensemble of lossy encoders. In essence, it is similar to the
one in \cite{DD11}, but there are a few technical differences, which we
use mainly for convenience.

For certain technical reasons that will become apparent later, we will
assume first that $|\calY|\ge |\calX|$ (and in Section 5, we will discuss the case where
this assumption is dropped).
Fix an arbitrarily small number
$\Delta > 0$. The codebook $\calC=\{\by_1,\ldots,\by_m\}$ is selected at random as
as follows: For each $\bx$ from a type class $\calT(Q_X)$ with
$H_Q(X)<\sqrt{\Delta}$, set the encoder output to be $\by\equiv\bx$, that
is, no distortion is incurred.\footnote{This distinction between
$H_Q(X)<\sqrt{\Delta}$ and $H_Q(X)\ge\sqrt{\Delta}$ is carried out for
technical reasons only, and it will be needed only in Section 6, where we
derive the compatible lower bound on the error probability of the ML decoder
(in other words, in Section 5, one can take $\Delta=0$).
In essence, for input sequences with very low empirical entropy, it makes
sense to apply lossless compression. This can only
improve the identification performance without compromising the compression
constraint.} 
For each type class with $H_Q(X)\ge\sqrt{\Delta}$ 
choose a certain conditional type
$Q_{Y|X}=\{Q_{Y|X}(y|x)~x\in\calX,~y\in\calY\}$ (depending on $Q_X$),
and then select uniformly at random
$M_Q=e^{nR_Q}$, $R_Q= I_Q(X;Y)+\Delta$
members of $\calT(Q_Y)$
to form a sub-code $\calC_Q=\{\by_\ell,~\ell=1,2,\ldots,M_Q\}$. 
The choice of $Q_{Y|X}$ is subjected to a compression constraint,
considering the fact that the compressed description of the encoder output is
of length approximately $nR_Q$ (plus an overhead of $O(\log n)$ bits that specify the type
$Q_X$). For example, to meet the expected length constraint,
$I_Q(X;Y)$ should 
not exceed $R_{\mbox{\tiny
C}}$ for all $Q_X$ in the vicinity of $G$.
For the excess length probability constraint, $I_Q(X;Y)$ must be kept less
than $R_{\mbox{\tiny C}}$ for every $Q_X$ with $D(Q_X\|G)\le E_{\mbox{\tiny
C}}$. For the exponential length moment constraint,
$sI_Q(X,Y)-D(Q_X\|G)$ must not exceed $\Gamma$ for any $Q_X$, namely,
$I_Q(X;Y)\le(\Lambda-D(Q_X\|G)/s$ for every $Q_X$.

For reasons that will become apparent later, we will 
assume that the choice of $Q_{Y|X}$,
for each $Q_X$, is such that the induced mapping $Q_X\to Q_Y$ is one--to--one,
namely, each $Q_Y$ is induced by no more than one $Q_X$.\footnote{
As a consequence of this fact, 
for $Q_X$ with $H_Q(X)<\sqrt{\Delta}$, we also have
$H_Q(Y)<\sqrt{\Delta}$. To maintain the one--to--one relation, it then
follows also that $H_Q(X)\ge \sqrt{\Delta}$ implies $H_Q(Y)\ge \sqrt{\Delta}$.} This means that 
given either $Q_X$ or $Q_Y$, the
entire joint type $Q_{XY}$ is fully determined.
Moreover, for technical reasons, we will assume that for each $Q_X$ with
$H_Q(X)\ge\sqrt{\Delta}$, $Q_{Y|X}$ is selected such that
$H_Q(X|Y)\ge\Delta+3\epsilon$, for some $0<\epsilon\ll\Delta$.
As said, each $\by_\ell\in\calC_Q$ is
selected independently at random 
under the uniform distribution within the type class 
of $Q_Y=\{Q_Y(y)~y\in\calY\}$,
where $Q_Y(y)=\sum_x Q_X(x)Q_{Y|X}(y|x)$.
The rate--distortion encoding rule is as follows. Each conditional type
$\calT(Q_{Y|X}|\bx)$, $\bx\in\calX^n$ 
(with $Q_{Y|X}$ matched to the type of $\bx$),
undergoes ranking according to a randomly chosen ordering of the members of
$\calT(Q_{Y|X}|\bx)$, under the uniform distribution across all
$|\calT(Q_{Y|X}|\bx)|!$ possible permutations.\footnote{The concept 
of ranking was already introduced
in the dual context, of channel decoding 
\cite{FL98}, \cite{LaZ98} as a convenient rule for resolving
ties.}
The orderings are independent for the various conditional types
$\{\calT(Q_{Y|X}|\bx),~\bx\in\calX^n\}$. Let $M(\bx,\by)$ denote the rank of
$\by\in\calT(Q_{Y|X}|\bx)$.
Let $\calM$ denote the set of randomly chosen ranking functions
$\{M(\bx,\by),~\bx\in\calX^n,~\by\in\calY^n\}$. 
Now, each $\bx_m\in\calT(Q_X)$ is encoded into the member of
$\calT_{Q_{Y|X}}(\bx_m)\bigcap\calC_Q$ with the smallest rank, $M(\bx_m,\by)$.
If $\calT_{Q_{Y|X}}(\bx_m)\bigcap\calC_Q=\emptyset$, the encoder outputs an arbitrary
$n$--tuple designating an error message (say, the all--zero sequence), 
without a hope for successful operation.
Let $f$ denote the resulting rate--distortion coding function, i.e., $\by=f(\bx)$.
The rate--distortion encoder
$f$ is therefore defined by the independent random selection
of both $\calC=\cup_QC_Q$ and $\calM$. 

\section{The Basic Idea}

The problem with the exact likelihood function (\ref{exactlikelihood}) is
that it is difficult to work with, both in the operative level, as an actual
decoding metric, and in the theoretical level, of
a single--letter performance analysis, and the reason, of course, is the multiplicative term
$\calI\{\bx\in f^{-1}(\by_m)\}$, that appears both in the numerator and the
denominator. Dasarathy and Draper \cite{DD11} have therefore analyzed
a simpler decoder --
the well known MMI decoder, which estimates $m$ according to the quantized
enrollment vector $\by_m$ with the highest value of $\hat{I}_{\by_m\bz}(Y;Z)$.
They have derived an achievable error exponent for a random selection of $f$,
which indicates that the MMI decoder is good enough to achieve the maximum rate
$R$ (channel capacity), given by $\max I(Y;Z)$, where the joint distribution of $(Y,Z)$ is
induced by a Markov chain $Y\to X\to Z$ and the maximization is over the
conditional distribution of $Y$ given $X$, which is subjected to a compression
constraint, $I(X;Y)\le R_{\mbox{\tiny C}}$, $R_{\mbox{\tiny C}}$ being the
allowed compression rate (see also \cite{Tuncel09}).

While the MMI decoder was shown to be sufficiently good to achieve capacity,
no further justification for this choice of decoder was
provided in \cite{DD11}.
A somewhat closer
inspection, however, reveals that there may be room for improvement in the choice
of the universal decoder, in order to achieve a better error exponent for a
given rate below capacity. This follows from the two following 
observations, which together form the basic idea of the paper.

The first observation is relevant even in the classical
random coding scenario, without the ingredient of lossy compression (i.e.,
$\by_m\equiv\bx_m$). Consider
then the ordinary random coding regime, where each codeword is selected
independently at random under the memoryless source $G$. Let the transmitted
codeword $\bx$ and the corresponding channel output $\bz$ be given. The
pairwise error event, that an independently generated competing codeword
$\bx^\prime$ would pose a threat to the correct decoding is lower bounded as
follows:
\begin{eqnarray}
\label{pairwiselb}
\sum_{\{\bx^\prime:~W(\bz|\bx^\prime)\ge W(\bz|\bx)\}}G(\bx^\prime)
&\ge& \sum_{\bx^\prime \in\calT(\bx|\bz)}G(\bx^\prime)\nonumber\\
&=& \sum_{\bx^\prime \in\calT(\bx|\bz)}G(\bx)\nonumber\\
&=& |\calT(\bx|\bz)|\cdot G(\bx)\nonumber\\
&\exe& \exp\{-n[\hI_{\bx\bz}(X;Z)+D(\hat{P}_{\bx}\|G)]\},
\end{eqnarray}
which is easily shown (using the method of types) to be achieved by the universal decoder $\hat{m}=\mbox{arg
max}_m[\hI_{\bx_m\bz}(X;Z)+D(\hat{P}_{\bx_m}\|G)]$ (see also \cite[eq.\
(16)]{givenclassofmetrics}). In other words, while the MMI decoding metric is
asymptotically optimal (in the random coding sense)
for the ensemble of {\it fixed composition codes}, when it comes to the
ensemble of i.i.d.\ random codewords, under $G$, this metric should be supplemented with
the divergence term, $D(\hat{P}_{\bx_m}\|G)$.

The second observation comes about when we put back the lossy compression
ingredient into our system model. In this case, the $\bx$--vectors in eq.\
(\ref{pairwiselb}) should be replaced by $\by$-vectors from the given codebook $\calC$, and the channel $W$
should be replaced by the channel $P$ defined in eq.\
(\ref{exactlikelihood}). Similarly, $G(\bx^\prime)$ should be replaced by
$P(\by^\prime)$, which is the denominator of (\ref{exactlikelihood}).
Suppose now that we can\footnote{This will indeed be shown later to be
possible for most encoders $\{f\}$ in the ensemble.}
approximate $P(\by^\prime)$ by
$e^{-n\alpha(\hat{P}_{\by^\prime})}$ (for $\by^\prime\in\calC$) and $P(\bz|\by)$ by
$e^{-n\beta(\hat{P}_{\by\bz})}$, where $\alpha(\cdot)$ and $\beta(\cdot)$ are
certain functions. Then, taking into account that $P(\by^\prime) > 0$ only for
$\by^\prime\in\calC$,
the analogue of the third line of (\ref{pairwiselb})
would now read
$|\calT(\by|\bz)\cap\calC|\cdot e^{-n\alpha(\hat{P}_{\by})}$,
a lower bound, which is asymptotically achieved by the universal decoder,
\begin{equation}
\label{univdec}
\hat{m}_{\mbox{\tiny u}}=\mbox{arg min}_m[\log N(\by_m|\bz)-n\alpha(\hat{P}_{\by_m})],
\end{equation}
where $N(\by|\bz)=|\calT(\by|\bz)\cap\calC|$, i.e., the number of codebook vectors
that are in the conditional type $\calT(\by|\bz)$.
In other words, our second observation is that in the 
problem setting considered here, the MMI decoder should be
modified, not only to account for the non--uniform input distribution, as
mentioned in the first observation above,
but also to account for the fact the support of this distribution is only
$\calC$, and not
$\calY^n$ in its entirety. In the next section, we will first specify the
function $\alpha(\cdot)$ and thereby fully define the proposed universal
decoder (\ref{univdec}).

\section{Main Result}

As mentioned in Section 3.2, since we assume that for each input assignment $Q_X$, the channel
$Q_{Y|X}$ is selected such that the mapping from $Q_X$ to $Q_Y=(Q_X\times
Q_{Y|X})_Y$ is one--one, a given $Q_Y$ can be induced from 
only one $Q_X$, which in turn dictates $Q_{Y|X}$, and hence
also the entire joint distribution $Q_{XY}$. In view of this, for a given
$Q_Y$ (or equivalently, a given $Q_{XY}$), let us define
\begin{equation}
A_Q(Y)=I_Q(X;Y)+D(Q_X\|G).
\end{equation}
To emphasize the dependence of $A_Q(Y)$ upon the empirical distribution of a
given $\by$, we also use the alternative notation $\alpha(\hP_{\by})$ instead of
$A_Q(Y)$, for every $\by\in\calT(Q_Y)$
(i.e., $\hP_{\by}=Q_Y$). Defining the universal decoder (\ref{univdec})
with this choice of the function $\alpha(\cdot)$,
we are now ready to state our main result.

\begin{theorem}
Consider the model and the assumptions described in Section 3 and the universal decoder
(\ref{univdec}) with the above definition of the function $\alpha(\cdot)$.
Then, for a given choice of $Q_{Y|X}$ as a functional of $Q_X$,
the random coding error exponent associated with the ensemble of codes,
described in Subsection 3.2, is given by
\begin{eqnarray}
\label{E(R)}
E(R_{\mbox{\tiny I}})&=&\min_{Q_X}\min_{Q_{Z|Y}}\left\{D(Q_X\|G)+
\min_{\tilde{Q}_{X|YZ}\in\calU(Q_{X|Y})}
D(\tilde{Q}_{XZ|Y}\|Q_{X|Y}\times W|Q_Y)+\right.\nonumber\\
& &\left.+\max\{[I_Q(Y;Z)-I_Q(X;Y)]_+,
[I_Q(Y;Z)+D(Q_X\|G)-R_{\mbox{\tiny I}}\}]_+\right\},
\end{eqnarray}
where, for a given $Q_{YZ}$, the set $\calU(Q_{X|Y})$ is defined to consist of
all conditional distributions $\{\tQ_{X|YZ}\}$ that are consistent with
$Q_{X|Y}$, that is, $\sum_{z\in\calZ}\tQ_{X|YZ}(x|y,z)Q_{Y|Z}(y|z)=Q_{X|Y}(x|y)$
for every $(x,y)\in\calX\times\calY$.
\end{theorem}
Before we prove this theorem, a brief discussion is in order.

First, observe that the objective function to be minimized in (\ref{E(R)}) is
a functional of $Q_X$ (or equivalently, $Q_{XY}$) and $Q_{YZ}$, or,
equivalently, $Q_{Z|Y}$, as $Q_Y$ is already dictated by $Q_X$. Since $Q_X$
and $Q_{Z|Y}$ are not subject to our control, they undergo minimization.
The controllable part is the choice of $Q_{Y|X}$, which is allowed to depend
on $Q_X$, but not on $Q_{Z|Y}$. Therefore, the expression of $E(R_{\mbox{\tiny
I}})$ should, in principle,
include also maximization over $Q_{Y|X}$ in between $\min_{Q_X}$ and $\min_{Q_{Z|Y}}$.
This maximization should be carried out, of course, subject to the compression
constraint, which limits $Q_{Y|X}$ to some subset denoted $\calQ$.
The caveat is, however, that there is no apparent guarantee that the
optimal $Q_{Y|X}$, as a functional of $Q_X$, would induce a one--to--one
mapping from $Q_X$ to $Q_Y$, a requirement that was already mentioned
in Subsection 3.2, and whose motivation will be explained
in the next paragraph.
Nonetheless, we show in the appendix (subsection A.1) that it is
possible to slightly modify the optimal $Q_{Y|X}$ by an arbitrarily small perturbation (and
thus lose an arbitrarily small amount from the optimal error exponent, due
to continuity) and thereby make the mapping $Q_X\to Q_Y$ one--to--one. It follows then that
we can approach arbitrarily closely the min--max--min expression,
\begin{eqnarray}
\label{minmaxmin}
& &\min_{Q_X}\max_{Q_{Y|X}\in\calQ}\min_{Q_{Z|Y}}\left\{D(Q_X\|G)+
\min_{\tilde{Q}_{X|YZ}\in\calU(Q_{X|Y})}
D(\tilde{Q}_{XZ|Y}\|Q_{X|Y}\times W|Q_Y)+\right.\nonumber\\
& &\left.+\max\{[I_Q(Y;Z)-I_Q(X;Y)]_+,
[I_Q(Y;Z)+D(Q_X\|G)-R_{\mbox{\tiny I}}]_+\}\right\}.
\end{eqnarray}

As promised in the previous paragraph (and earlier), 
we now explain the motivation for insisting on a one--to--one
mapping $Q_X\to Q_Y$. The easiest way to see this is to look at the expression
$|\calT(\by|\bz)\cap\calC|\cdot \exp\{-n\alpha(\hP_{\by})\}$, which appears in the
last paragraph of Section 3, in the context of an achievable lower bound to
the pairwise error probability for a given $(\by,\bz)$. We would like, of
course, to keep
this quantity as small as possible. Now, in general, if $Q_X\to Q_Y$ is not
necessarily one--to--one, $\calT(\by|\bz)\cap\calC$ may include reproduction vectors that
correspond to $\bx$--vectors from all types $\{Q_X\}$ that are mapped to
the given $Q_Y=\hP_{\by}$, but if $Q_X\to Q_Y$ is one--to--one, then there is only one
such $Q_X$. Moreover, a many--to--one relation $Q_X\to Q_Y$ may decrease the
above exponential term $\alpha(\hP_{\by})$ (i.e., increase 
the factor $\exp\{-n\alpha(\hP_{\by})\}$) since the given $\by$ may have more
types $\{Q_X\}$ of source
vectors $\{\bx\}$ that could yield the given $\by$ using the source encoder.
In particular, the definition of $A_Q(Y)$ should then include also a minimization
over all $\{Q_{X|Y}\}$ pertaining to $\{Q_X\}$ that are mapped to the given
$Q_Y$, which may again result in degradation in performance. 
But when $Q_X\to Q_Y$ is one--to-one, as required, there is only one such $Q_X$.
More precisely, in view of the above discussion, it is possible to show that if
the requirement of a one--to--one mapping $Q_X\to Q_Y$ is dropped (and
then there is no longer need to assume $|\calY|\ge|\calX$, and we can also take
$\Delta=0$), then the
term in the second line of (\ref{minmaxmin}) should be replaced by the
following expression:
\begin{eqnarray}
\label{not1-1}
& &\left[I_Q(Y;Z)-\max_{\tilde{Q}_{X|Y}\in\calS(Q_Y)}I_{\tilde{Q}}(X;Y)\right]_++
\left[\min_{\tilde{Q}_{X|Y}\in\calS(Q_Y)}
\{I_{\tilde{Q}}(X;Y)+D(\tilde{Q}_X\|G)\}-\right.\nonumber\\
& &\left.\left[\max_{\tilde{Q}_{X|Y}\in\calS(Q_Y)}
I_{\tilde{Q}}(X;Y)-I_Q(Y;Z)\right]_+-R_{\mbox{\tiny
I}}\right]_+,
\end{eqnarray}
where $\calS(Q_Y)$ is the collection of all $\tilde{Q}_{X|Y}$ such that
$\tilde{Q}_X=(Q_Y\times \tilde{Q}_{X|Y})_X$ is mapped to $Q_Y$. Clearly, the
larger is the set $\calS(Q_Y)$, the smaller is the resulting expression,
and so, the best one can hope for is that $\calS(Q_Y)$ would be a singleton,
in which case, it becomes identical to the term in the second line of (\ref{minmaxmin}).
Nonetheless, it should be pointed out that even in the general case, where
$Q_X\to Q_Y$ is not one--to--one, 
and hence $\calS(Q_Y)$ is not a singleton,
the resulting error exponent cannot be worse than that of \cite{DD11}, since
our proposed universal decoder is at least as good as any other decoder whose metric
depends only on the empirical joint distribution of $(\by_m,\bz)$ (see item 4
in the Introduction) and in particular, it is also as good as the ML
decoder (see Section 6). Here, we should remark that the modification
(\ref{not1-1}) significantly complicates the optimization of $Q_{Y|X}$ for a
given $Q_X$, because (\ref{not1-1}) depends on the mapping
$Q_X=U[Q_{Y|X}]$ in a {\it global} manner (via the sets $\calS(Q_Y)$, induced by
$U[\cdot]$) and
not only in a local, pointwise manner, 
of optimizing $Q_{Y|X}$ for each given $Q_X$ separately. 
Therefore, the appropriate way to present the
error exponent expression, in this more general case, is in terms of 
the series of optimizations, $\sup_{U[\cdot]}\min_{Q_X}\min_{Q_{X|Y}}$,
rather than the min--max--min as before. (Of course, the supremum over $U[\cdot]$
is subject to the compression constraint.)

Finally, a word on the comparison between our result (\ref{minmaxmin}) and the one
in \cite[Theorem 1]{DD11}, is in order. The first two terms in (\ref{minmaxmin}) are
identical to those in \cite[Theorem 1]{DD11}, as they are just the terms of the
exponential probabilistic weighting 
of the dominant type $Q_{YZ}$, i.e., the one that contributes most to
the probability of error. However, the third term in
(\ref{minmaxmin}) is different from the one in \cite{DD11}, which, in our
notation, is simply $[I_Q(Y;Z)-R_{\mbox{\tiny I}}]_+$. Even if we 
ignore the term $[I_Q(Y;Z)-I_Q(X;Y)]_+$ in the second line of
(\ref{minmaxmin}), and lower bound
our third term just by $[I_Q(Y;Z)+D(Q_X\|G)-R_{\mbox{\tiny I}}]_+$, it
obviously cannot be smaller than $[I_Q(Y;Z)-R_{\mbox{\tiny I}}]_+$, of
\cite{DD11}, due to the divergence term, $D(Q_X\|G)$. It is clear then that,
at least at low rates (say, even $R_{\mbox{\tiny I}}=0$), the exponent
(\ref{minmaxmin}) is strictly larger than that of \cite{DD11} whenever the
minimizing $Q_X$ differs from $G$, which can indeed be the case in many
situations (see Subsection A.2 of the appendix for a demonstration of this
fact).\\

\noindent
{\it Proof of Theorem 1.}
We begin with a simple upper bound to $P(\by)$ for $\by\in\calC\in\calT(Q_Y)$, which
applies to every $f=(\calC,\calM)$ since $f^{-1}(\by)\subseteq\calT(Q_{X|Y}|\by)$,
where $Q_{X|Y}$ is the reverse channel that corresponds to $Q_Y$:
\begin{eqnarray}
P(\by)&=&\sum_{\bx\in\calX^n}G(\bx)\calI\{\bx\in f^{-1}(\by)\}\\
&\le&|\calT(Q_{X|Y}|\by)|\cdot G(\bx)\\
&\le&\exp\{nH_Q(X|Y)+O_1(\log n)\}\cdot \exp\{-n[H_Q(X)+D(Q_X\|G)]\}\\
&=&\exp\{-n[I_Q(X;Y)+D(Q_X\|G)+O_1(\log n)]\}\\
&=&e^{-nA_Q(Y)+O_1(\log n)},
\end{eqnarray}
where $O_1(\log n)$ is a quantity (resulting from the method of types), whose leading 
term is proportional to $\log n$.
Similarly, for $(\by,\bz)\in\calT(Q_{YZ})$ with $\by\in \calC\in\calT(Q_Y)$,
we have
\begin{eqnarray}
P(\by,\bz)&=&\sum_{\bx\in\calX^n}G(\bx)W(\bz|\bx)\calI\{\bx\in f^{-1}(\by)\}\\
&\le&\sum_{\{T(Q_{X|YZ}|\by,\bz):~Q_{X|YZ}\in\calU(Q_{X|Y})\}}|\calT(Q_{X|YZ}|\by.\bz)|\cdot
\left[G(\bx)W(\bz|\bx)\right]_{(\bx,\bz)\in\calT(Q_{XZ})}\\
&\le&\max_{Q_{X|YZ}\in\calU(Q_{X|Y})}\exp\{nH_Q(X|Y,Z)+O_2(\log
n)\}\times\nonumber\\
& &\exp\{n\sum_{x,z}Q_{XZ}(x,z)\log[G(x)W(z|x)\}\\
&=&\exp\left\{-n\min_{Q_{X|YZ}\in\calU(Q_{X|Y})}\sum_{x,y,z}Q_{XYZ}(x,y,z)
\log\frac{Q_{X|YZ}(x|y,z)}{G(x)W(z|x)}+O_2(\log n)
\right\}\\
&\dfn& e^{-nB_Q(Y,Z)+O_2(\log n)},
\end{eqnarray}
where $O_2(\log n)$ is again a quantity dominated by a term proportional to
$\log n$.
For later use, the following algebraic manipulation will be found useful.
\begin{eqnarray}
\label{bqyz}
B_Q(Y,Z)&=&\min_{Q_{X|YZ}\in\calU(Q_{X|Y},0)}\sum_{x,y,z}Q_{XYZ}(x,y,z)
\log\frac{Q_{X|YZ}(x|y,z)}{G(x)W(z|x)}\nonumber\\
&=&\min_{Q_{X|YZ}\in\calU(Q_{X|Y})}[H_Q(X,Z)+D(Q_{XZ}\|G\times
W)-H_Q(X|Y,Z)]\nonumber\\
&=&\min_{Q_{X|YZ}\in\calU(Q_{X|Y})}[H_Q(X)+H_Q(Z|X)-H_Q(X|Y,Z)+D(Q_{XZ}\|G\times
W)]\nonumber\\
&=&\min_{Q_{X|YZ}\in\calU(Q_{X|Y})}[I_Q(X;Y,Z)+H_Q(Z|X)+D(Q_{XZ}\|G\times
W)]\nonumber\\
&=&\min_{Q_{X|YZ}\in\calU(Q_{X|Y})}[H_Q(Y,Z)-H_Q(Y,Z|X)+H_Q(Z|X)+D(Q_{XZ}\|G\times
W)]\nonumber\\
&=&H_Q(Y,Z)+\min_{Q_{X|YZ}\in\calU(Q_{X|Y})}[D(Q_{XZ}\|G\times
W)-H_Q(Y|X,Z)].
\end{eqnarray}
Now, consider the universal decoding metric
\begin{equation}
d(\by,\bz)=\log N(\by|\bz)-n\alpha(\hP_{\by}).
\end{equation}
Then, defining $\calE_{\mbox{\tiny u}}(\by,\bz)=\{\by':~d(\by',\bz)\le
d(\by,\bz)\}\cap\calC$, we have
\begin{eqnarray}
\sum_{\by'\in \calE_{\mbox{\tiny u}}(\by,\bz)}
P(\by')&=&\sum_{\{\calT(\by'|\bz):~\calT(\by'|\bz)\cap\calC\subseteq 
\calE_{\mbox{\tiny u}}(\by,\bz)\}} P[\calC\cap\calT(\by'|\bz)]\\
&=&\sum_{\{\calT(\by'|\bz):~\calT(\by'|\bz)\cap\calC\subseteq 
\calE_{\mbox{\tiny u}}(\by,\bz)\}} |\calC\cap\calT(\by'|\bz)|\cdot P(\by')\\
&\lexe&\sum_{\{\calT(\by'|\bz):~\calT(\by'|\bz)\cap\calC\subseteq 
\calE_{\mbox{\tiny u}}(\by,\bz)\}} N(\by'|\bz)\cdot e^{-n\alpha(\hP_{\by'})}\\
&\le&\sum_{\{\calT(\by'|\bz):~\calT(\by'|\bz)\cap\calC\subseteq 
\calE_{\mbox{\tiny u}}(\by,\bz)\}} N(\by|\bz)\cdot e^{-n\alpha(\hP_{\by})}\\
&\exe& N(\by|\bz)\cdot e^{-n\alpha(\hP_{\by})}.
\end{eqnarray}
Then, for a given $f=(\calC,\calM)$, the probability of error
of the universal decoder (\ref{univdec}),
$\mbox{P}_{\mbox{\tiny e,u}}(f)$, is upper bounded as follows.
\begin{eqnarray}
\label{ubpe}
\mbox{P}_{\mbox{\tiny
e,u}}(f)&=&\sum_{\by\in\calC}\sum_{\bz\in\calZ^n}P(\by,\bz)\cdot\min\left\{1,e^{nR_{\mbox{\tiny
I}}}\cdot \sum_{\by'\in\calE_{\mbox{\tiny
u}}(\by,\bz)}P(\by')\right\}\nonumber\\
&\lexe&\sum_{\by\in\calC}\sum_{\bz\in\calZ^n}P(\by,\bz)\cdot\min\left\{1,e^{n[R_{\mbox{\tiny
I}}-\alpha(\hP_{\by})]}\cdot N(\by|\bz)\right\}\nonumber\\
&=&\sum_{\by\in\calY^n}\calI\{\by\in\calC\}\cdot
\sum_{\bz\in\calZ^n}P(\by,\bz)\cdot\min\left\{1,e^{n[R_{\mbox{\tiny
I}}-\alpha(\hP_{\by})]}\cdot N(\by|\bz)\right\}.
\end{eqnarray}
From this point onward, we will average the upper bound on
$\mbox{P}_{\mbox{\tiny e,u}}(f)$ across the ensemble of $\{f\}$.
This will be done in two steps. In the first step, we average over all
incorrect codewords, whose contributions are expressed in the random variable
$N(\by|\bz)$. In the second step, we average over the correct codeword (which
is drawn independently of all incorrect codewords), that
is expressed in the factor $\calI\{\by\in\calC\}$ in the last expression.
Now, for a given pair 
$(\by,\bz)\in\calT(Q_{YZ})$, the number $N(\by|\bz)$ is a binomial random
variable (RV) with
$e^{n[I_Q(X;Y)+\Delta]}$ trials
and probability of success of the exponential 
order of $e^{-nI_Q(Y;Z)}$. Thus, for a given $\epsilon > 0$, if
$I_Q(X;Y)+\Delta\ge
I_Q(Y;Z)$, then
$N(\by|\bz)\le e^{n[I_Q(X;Y)-I_Q(Y;Z)+\Delta+\epsilon]}$ with probability at least as larger as
$1-\exp[-(n\epsilon-1)e^{n\epsilon}]$ (as can easily been seen from a
derivation similar to the one in \cite[pp.\ 167--168]{fnt}).
For $I_Q(X;Y)+\Delta< I_Q(Y;Z)$, the RV $N(\by|\bz)$ 
exceeds unity with
probability of the exponential order of $e^{-n[I_Q(Y;Z)-I_Q(X;Y)-\Delta]}$ 
(similarly to \cite[eq.\ (6.36)]{fnt}) and
it exceeds the value $e^{n\epsilon}$, with probability less than
$\exp[-(n\epsilon-1)e^{n\epsilon}]$.
It follows then that for a given deterministic $s$, and for
$I_Q(X;Y)+\Delta\ge
I_Q(Y;Z)$, 
\begin{eqnarray}
\bE\left[\min\left\{1,e^{-ns}N(\by|\bz)\right\}\right]&\lexe& 
\min\left\{1,e^{-ns}\cdot e^{n[I_Q(X;Y)-I_Q(Y;Z)+\Delta+\epsilon]}\right\}\\
&\exe&
\exp\{-n[s+I_Q(Y;Z)-I_Q(X;Y)-\Delta-\epsilon]_+\},
\end{eqnarray}
whereas for $I_Q(X;Y)+\Delta<
I_Q(Y;Z)$, 
\begin{eqnarray}
\bE\left[\min\left\{1,e^{-ns}N(\by|\bz)\right\}\right]&\lexe& 
e^{-n[I_Q(Y;Z)-I_Q(X;Y)-\Delta]}\cdot\min\left\{1,e^{-ns}\right\}\\
&=&
\exp\{-n[I_Q(Y;Z)-I_Q(X;Y)-\Delta+[s]_+]\}.
\end{eqnarray}
Since we are interested merely in the exponential order, from now on, we shall neglect the
$\Delta$ and $\epsilon$ terms, which eventually tends to zero anyway. The last two equations
can now be unified as follows:
\begin{eqnarray}
& &\bE\left[\min\left\{1,e^{-ns}N(\by|\bz)\right\}\right]\nonumber\\
&\lexe&
\exp\left\{-n\left([I_Q(Y;Z)-I_Q(X;Y)]_++[s-[I_Q(X;Y)-I_Q(Y;Z)]_+]_+\right)\right\}.
\end{eqnarray}
This exponential upper bound will be applied with the assignment
$s=\alpha(\hP_{\by})-R_{\mbox{\tiny I}}$ (or equivalently,
$s=A_Q(Y)-R_{\mbox{\tiny I}}$). As for averaging over the randomness of the correct
codeword, note that for a given $\by\in\calT(Q_Y)$,
\begin{equation}
\bE[\calI\{\by\in\calC\}]=\mbox{Pr}\{\by\in\calC\}=
1-\left(1-\frac{1}{|\calT(Q_Y)|}\right)^{e^{n[I_Q(X;Y)+\Delta]}}\exe
e^{-n[H_Q(Y|X)-\Delta]}.
\end{equation}
Putting all this altogether, we obtain (again, neglecting $\Delta$):
\begin{eqnarray}
\label{almostfinal}
\bar{\mbox{P}}_{\mbox{\tiny
e,u}}&\dfn&\bE\left\{\mbox{P}_{\mbox{\tiny
e}}(f)\right\}\nonumber\\
&\lexe&\sum_{Q_Y}|\calT(Q_Y)|\cdot 
e^{-nH_Q(Y|X)}\sum_{Q_{Z|Y}}|\calT(Q_{Z|Y})|\cdot e^{-nB_Q(Y,Z)}\times\nonumber\\
& &\exp\left\{-n\left([I_Q(Y;Z)-I_Q(X;Y)]_++[A_Q(Y)-[I_Q(X;Y)-I_Q(Y;Z)]_+-R_{\mbox{\tiny
I}}]_+\right)\right\}\nonumber\\
&\exe&\exp\left\{-n\min_{Q_{YZ}}\left(B_Q(Y,Z)-H_Q(Z|Y)-H_Q(Y)+H_Q(Y|X)
+\right.\right.\nonumber\\
& &\left.\left.[I_Q(Y;Z)-I_Q(X;Y)]_+
+[A_Q(Y)-[I_Q(X;Y)-I_Q(Y;Z)]_+-R_{\mbox{\tiny I}}]_+\right)\right\}\nonumber\\
&=&\exp\left\{-n\min_{Q_{YZ}}\left(B_Q(Y,Z)-H_Q(Y,Z)+H_Q(Y|X)
+[I_Q(Y;Z)-I_Q(X;Y)]_++\right.\right.\nonumber\\
& &\left.\left.+[I_Q(X;Y)+D(Q_X\|G)-[I_Q(X;Y)-I_Q(Y;Z)]_+-R_{\mbox{\tiny
I}}]_+\right)\right\}\nonumber\\
&=&\exp\left\{-n\min_{Q_X,Q_{Z|Y}}\left(B_Q(Y,Z)-H_Q(Y,Z)+H_Q(Y|X)+
[I_Q(Y;Z)-I_Q(X;Y)]_++\right.\right.\nonumber\\
& &\left.\left.[I_Q(X;Y)+D(Q_X\|G)-[I_Q(X;Y)-I_Q(Y;Z)]_+
-R_{\mbox{\tiny I}}]_+\right)\right\}.
\end{eqnarray}
To simplify the above expression, 
and to modify its form to one that is more easily comparable to \cite{DD11},
we first observe (using (\ref{bqyz})) that
\begin{eqnarray}
& &B_Q(Y,Z)-H_Q(Y,Z)+H_Q(Y|X)\nonumber\\
&=&\min_{\tilde{Q}_{X|YZ}\in\calU(Q_{X|Y})}[
D(Q_{XZ}\|G\times
W)-H_Q(Y|X,Z)]+H_Q(Y|X)\\
&=&\min_{\tilde{Q}_{X|YZ}\in\calU(Q_{X|Y})}[D(Q_{XZ}\|G\times
W)+I_{\tilde{Q}}(Y;Z|X)]\\ 
&=&D(Q_X\|G)+\min_{\tilde{Q}_{X|YZ}\in\calU(Q_{X|Y})}[D(\tilde{Q}_{Z|X}\|W|Q_X)+
I_{\tilde{Q}}(Y;Z|X)]\\
&=&D(Q_X\|G)+\min_{\tilde{Q}_{X|YZ}\in\calU(Q_{X|Y})}\sum_{y,z}Q_{YZ}(y,z)\times\\
& &\sum_x
\tilde{Q}_{X|YZ}(x|y,z)\log\left[\frac{\tilde{Q}_{Z|X}(z|x)}{W(z|x)}
\cdot\frac{\tilde{Q}_{YZ|X}(y,z|x)}
{\tilde{Q}_{Z|X}(z|x)Q_{Y|X}(y|x)}\right]\\
&=&D(Q_X\|G)+\min_{\tilde{Q}_{X|YZ}\in\calU(Q_{X|Y})}\sum_{y,z}Q_{YZ}(y,z)\times\\
& &\sum_x
\tilde{Q}_{X|YZ}(x|y,z)\log\left[\frac{\tilde{Q}_{YZ|X}(y,z|x)}
{W(z|x)Q_{Y|X}(y|x)}\right]\\
&=&D(Q_X\|G)+\min_{\tilde{Q}_{X|YZ}\in\calU(Q_{X|Y})}\sum_{y,z}Q_{YZ}(y,z)\times\\
& &\sum_x
\tilde{Q}_{X|YZ}(x|y,z)\log\left[\frac{\tilde{Q}_{Z|XY}(z|x,y)}
{W(z|x)}\right]\\
&=&D(Q_X\|G)+\min_{\tilde{Q}_{X|YZ}\in\calU(Q_{X|Y})}\sum_{y,z}Q_{YZ}(y,z)\times\\
& &\sum_x
\tilde{Q}_{X|YZ}(x|y,z)\log\left[\frac{\tilde{Q}_{Z|XY}(z|x,y)Q_{X|Y}(x|y)}
{W(z|x)Q_{X|Y}(x|y)}\right]\\
&=&D(Q_X\|G)+\min_{\tilde{Q}_{X|YZ}\in\calU(Q_{X|Y})}\sum_{y,z}Q_{YZ}(y,z)\times\\
& &\sum_x
\tilde{Q}_{X|YZ}(x|y,z)\log\left[\frac{\tilde{Q}_{XZ|Y}(x,z|y)}
{W(z|x)Q_{X|Y}(x|y)}\right]\\
&=&D(Q_X\|G)+\min_{\tilde{Q}_{X|YZ}\in\calU(Q_{X|Y})}
D(\tilde{Q}_{XZ|Y}\|Q_{X|Y}\times W|Q_Y),
\end{eqnarray}
which are the first two terms in (\ref{E(R)}).
As for the other terms of (\ref{almostfinal}), we use the identities
$a-[a-b]_+\equiv b-[b-a]_+\equiv\min\{a,b\}$ and $b+[a-b]_+\equiv\max\{a,b\}$
to obtain
\begin{eqnarray}
& &[I_Q(Y;Z)-I_Q(X;Y)]_+
+[I_Q(X;Y)+D(Q_X\|G)-[I_Q(X;Y)-I_Q(Y;Z)]_+-R_{\mbox{\tiny I}}]_+\\
&=&[I_Q(Y;Z)-I_Q(X;Y)]_+
+[I_Q(Y;Z)+D(Q_X\|G)-[I_Q(Y;Z)-I_Q(X;Y)]_+-R_{\mbox{\tiny I}}]_+\\
&=&\max\{[I_Q(Y;Z)-I_Q(X;Y)]_+,I_Q(Y;Z)+D(Q_X\|G)-R_{\mbox{\tiny I}}\}\\
&=&\max\{[I_Q(Y;Z)-I_Q(X;Y)]_+,[I_Q(Y;Z)+D(Q_X\|G)-R_{\mbox{\tiny I}}]_+\},
\end{eqnarray}
which is the last term in (\ref{E(R)}).
This completes the proof of Theorem 1.

\section{A Matching Lower Bound on ML Decoding Performance}

In this section, we argue that the proposed universal decoder is
asymptotically optimal in the sense that its error exponent is the same
as that of the ML decoder, at least for channels with strictly positive
single--letter transition probabilities, $\{W(z|x)\}$. The limitation
to strictly positive $\{W(z|x)\}$ is rather technical, but it is conjectured
that this argument continues to hold true even without this restriction. The
reason for this belief is that random coding error exponents are normally
continuous functionals of the channel parameters, and therefore, it is
seems inconceivable that there would be significant differences between the
error exponent of a channel where some $\{W(z|x)\}$ vanish and the one of a
nearby channel where the parameters are slightly altered so that all $\{W(z|x)\}$
are positive.

\begin{theorem}
Let $W$ be a DMC with strictly positive single--letter probabilities,
$\{W(z|x)\}$ and
consider the model described in Section 3 along with the ML decoder, based on
(\ref{exactlikelihood}).
Then, for a given choice of $Q_{Y|X}$ as a functional of $Q_X$,
the random coding error exponent associated with the ensemble of codes,
described in Subsection 3.2 and ML decoding, is given by eq.\ (\ref{E(R)}).
\end{theorem}

\noindent
{\it Proof of Theorem 2.} Since the ML decoder cannot be worse than the
universal decoder (\ref{univdec}), it is enough to prove that average error
probability of the ML decoder is lower bounded by an expression of the
exponential order of $e^{-nE(R_{\mbox{\tiny I}})}$. The analysis is basically
with the same
method as in the proof of Theorem 1, except that here, we are after lower
bounds (rather than upper bounds) to certain expressions. 

We begin with
lower bounds on $P(\by)$ and $P(\by,\bz)$, but to this end, we first need some
preparatory steps.
For a given $\bx\in\calT(Q_X)$ and $\by\in\calC_Q\cap\calT(Q_{Y|X}|\bx)$, 
we first observe that
\begin{equation}
\calI\{\bx\in f^{-1}(\by)\}=\prod_{\by'\in\calC_Q\cap\calT(Q_{Y|X}|\bx)}
[1-\calI\{M(\bx,\by')<
M(\bx,\by)\}].
\end{equation}
Due to the symmetry of the random selection of $\calM$, it
is clear that for a given $\bx\in\calT(Q_X)$ and $\calC_Q$, 
every $\by\in\calC_Q\cap\calT(Q_{Y|X}|\bx)$
has exactly the same probability to have the smallest rank among all members of
$\calC_Q\cap\calT(Q_{Y|X}|\bx)$, and so, this probability is
$1/|\calC_Q\cap\calT(Q_{Y|X}|\bx)|$. 
Next observe that $|\calC_Q\cap\calT(Q_{Y|X}|\bx)|$ is a
binomial RV with $|\calC_Q|=e^{n[I_Q(X;Y)+\Delta]}$ trials and probability of
success of the exponential order of $e^{-nI_Q(X;Y)}$, 
therefore $|\calC_Q\cap\calT(Q_{Y|X}|\bx)|$
concentrates double--exponentially rapidly around $e^{n\Delta}$. In fact, this
is true for
the vast majority of rate--distortion codes.
More precisely, let $0 < \epsilon \ll \Delta$ be given. 
Then, for every given $Q_X$ with $H_Q(X)\ge\sqrt{\Delta}$, its associated $Q_{Y|X}$, 
and $\bx\in\calT(Q_X)$,
\begin{equation}
\label{de1}
\mbox{Pr}\left\{|\calC_Q\cap\calT(Q_{Y|X}|\bx)| \ge
e^{n(\Delta+\epsilon)}\right\}\le
\exp\left\{-(n\epsilon-1)e^{n\Delta}\right\}
\end{equation}
and
\begin{equation}
\label{de2}
\mbox{Pr}\left\{|\calC_Q\cap\calT(Q_{Y|X}|\bx)| \le
e^{n(\Delta-\epsilon)}\right\}\le
\exp\left\{-[1-(n\epsilon+1)e^{-n\epsilon}]e^{n\Delta}\right\}. 
\end{equation}
From now on, suppose that $\calC$ belongs to the vast majority of codes that
satisfy
\begin{equation}
\label{goodc}
e^{n(\Delta-\epsilon)}\le
|\calC_Q\cap\calT(Q_{Y|X}|\bx)|\le e^{n(\Delta+\epsilon)}~~~~\forall
\bx\in\cup_{Q_X:~H_Q(X)\ge\sqrt{\Delta}}\calT(Q_X).
\end{equation}
Next, for a given $\calC=\cup_Q\calC_Q$, 
since the various random ordering functions
$\{M(\bx,\cdot),~\bx\in\calT(Q_{X|Y}|\by)\}$ are independent, the quantity
$|\calT(Q_{X|Y}|\by)\cap f^{-1}(\by)|$ is a binomial RV with exponentially
$e^{nH_Q(X|Y)}$ trials and probability of success
$1/|\calC_Q\cap\calT(Q_{Y|X}|\bx)|\exe e^{-n(\Delta\pm\epsilon)}$. 
Therefore, since $H_Q(X|Y)$ is assumed at least as large as $\Delta+3\epsilon$
whenever $H_Q(X)\ge\sqrt{\Delta}$ (by the code construction described in Section 3.2), then 
\begin{equation}
\label{de3}
\mbox{Pr}\left\{|\calT(Q_{X|Y}|\by)\cap f^{-1}(\by)|\le 
e^{n[H_Q(X|Y)-\Delta-2\epsilon]}\right\}\le
\exp\left\{-e^{n\epsilon}+n\epsilon+1\right\}.
\end{equation}
Let us define now the class $\calG$ of codes $f=(\calC,\calM)$ that
satisfy (\ref{goodc}) as well as the following two conditions. The first
condition is that
\begin{equation}
\label{goodm1}
|\calT(Q_{X|Y}|\by)\cap f^{-1}(\by)|\ge
e^{n[H_Q(X|Y)-\Delta-2\epsilon]} 
\end{equation}
for every $\by\in\calC\cap\calT(Q_Y)$ with $H_Q(Y)\ge\sqrt{\Delta}$, and
the second condition is that
\begin{equation}
\label{goodm2}
|\calT(Q_{X|YZ}|\by,\bz)\cap f^{-1}(\by)|\ge
e^{n[H_Q(X|YZ)-\Delta-2\epsilon]} 
\end{equation}
for every $(\by,\bz)\in\calT(Q_{YZ})$ such that $\by\in\calC$,
$H_Q(Y)\ge\sqrt{\Delta}$, and 
with $Q_{X|YZ}$
such that $H_Q(X|Y,Z)\ge\Delta+3\epsilon$. 
The double--exponential decay of the probabilities (\ref{de1}), (\ref{de2}) and
(\ref{de3}) imply that the vast majority of codes $f=(\calC,\calM)$ are in
$\calG$, in particular, $\calG$ contains a fraction of the codes 
that tends to one double--exponentially.

Consider an arbitrary code $f=(\calC,\calM)\in\calG$,
and let $\by\in\calC\cap\calT(Q_Y)$ be given.
Obviously, for $Q_Y$ with $H_Q(Y)<\sqrt{\Delta}$,
$P(\by)=G(\by)=\exp\{-n[H_Q(Y)+D(Q_Y\|G)]\}$ since
$\by\equiv\bx$. For $H_Q(Y)\ge\sqrt{\Delta}$,
since $|\calC_Q\cap \calT(Q_{Y|X}|\bx)|\ge e^{n(\Delta-\epsilon)}$, we
have
\begin{eqnarray}
P(\by)&=&\sum_{\bx}G(\bx)\calI\{\bx\in
f^{-1}(\by)\}\\
&=&G(\bx)\bigg|_{\bx\in\calT(Q_X)}\cdot|\calT(Q_{X|Y}|\by)\cap
f^{-1}(\by)|\\
&\ge&
\exp\{-n[H_Q(X)+D(Q_X\|G)-H_Q(X|Y)+\Delta+2\epsilon]-O_1(\log n)\}\\
&=&\exp\{-n[A_Q(Y)+\Delta+2\epsilon]-O_1(\log n)\}.
\end{eqnarray}
Note that this lower bound to $P(\by)$ applies also to $Q_Y$ with
$H_Q(Y)<\sqrt{\Delta}$, where $X\equiv Y$, since 
$A_Q(Y,Y)=I_Q(Y;Y)+D(Q_Y\|G)=H_Q(Y)+D(Q_Y\|G)$.

Next, consider a pair $(\by,\bz)\in\calT(Q_{YZ})$ with $\by\in\calC$.
Again, if $H_Q(Y)<\sqrt{\Delta}$,
\begin{equation}
P(\by,\bz)=G(\by)W(\bz|\by)=\exp\{-n[H_Q(Y,Z)+D(Q_{YZ}\|G\times W)]\}.
\end{equation}
For $H_Q(Y)\ge\sqrt{\Delta}$ (and hence also $H_Q(X)\ge\sqrt{\Delta}$),
define the set
\begin{eqnarray}
\calU(Q_{X|Y},\Delta)&=&\{Q_{X|YZ}:~H_Q(X|Y,Z)\ge\Delta,~\nonumber\\
& &\sum_zQ_{Z|Y}(z|y)Q_{X|YZ}(x|y,z)=Q_{X|Y}(x|y),~\forall~x,y\},
\end{eqnarray}
where, of course, $\calU(Q_{X|Y},0)$ is identical to $\calU(Q_{X|Y})$ defined
before. Then, for $f\in\calG$,
\begin{eqnarray}
P(\by,\bz)&=&\sum_{\bx\in\calX^n}G(\bx)W(\bz|\bx)\calI\{\bx\in f^{-1}(\by)\}\\
&=&\sum_{\calT(Q_{X|YZ}|\by,\bz):~Q_{X|YZ}\in\calU(Q_{X|Y},0)}
[G(\bx)W(\bz|\bx)]\bigg|_{(\bx,\bz)\in\calT(Q_{XZ})}\times\nonumber\\
& &\sum_{\bx\in
\calT(Q_{X|YZ}|\by,\bz)}\calI\{\bx\in f^{-1}(\by)\}\\
&\ge&\sum_{\{\calT(Q_{X|YZ}|\by,\bz):~Q_{X|YZ}\in\calU(Q_{X|Y},\Delta+3\epsilon)\}} 
[G(\bx)W(\bz|\bx)]\bigg|_{(\bx,\bz)\in\calT(Q_{XZ})}\times\nonumber\\
& &\sum_{\bx\in
\calT(Q_{X|YZ}|\by,\bz)}\calI\{\bx\in f^{-1}(\by)\}\\
&\ge&\sum_{\{\calT(Q_{X|YZ}|\by,\bz):~Q_{X|YZ}\in\calU(Q_{X|Y},\Delta+3\epsilon)\}}
[G(\bx)W(\bz|\bx)]\bigg|_{(\bx,\bz)\in\calT(Q_{XZ})}\times\nonumber\\
& &\exp\{n[H_Q(X|Y,Z)-\Delta-2\epsilon]-O_2(\log n)\}\\
&\ge&e^{-n(\Delta+2\epsilon)-O_2(\log n)}\times\nonumber\\
& &\exp\left\{-n\min_{Q_{X|YZ}
\in\calU(Q_{X|Y},\Delta+3\epsilon)}\sum_{x,y,z}Q_{XYZ}(x,y,z)
\log\frac{Q_{X|YZ}(x|y,z)}{G(x)W(z|x)}\right\}\\
&\ge&\exp\left\{-n\left[\Delta+2\epsilon+
\left(\sqrt{\Delta}+\frac{3\epsilon}{\sqrt{\Delta}}\right)
\max_{x,z}\log\frac{1}{G(x)W(z|x)}\right]-O_2(\log n)\right\}\times\nonumber\\
& &\exp\left\{-n\min_{Q_{X|YZ}\in\calU(Q_{X|Y})}\sum_{x,y,z}Q_{XYZ}(x,y,z)
\log\frac{Q_{X|YZ}(x|y,z)}{G(x)W(z|x)}\right\}\\
&\dfn&\exp\{-n\theta(\Delta,\epsilon)-O_2(\log n)\}\cdot
\exp\{-nB_Q(Y,Z)\}, 
\end{eqnarray}
where $\lim_{\Delta\to 0}\lim_{\epsilon\to 0}\theta(\Delta,\epsilon)=0$,
provided that $W(z|x)$ for every $(x,z)$,
and where the second to the last step follows from the following consideration.
Let $Q_{X|YZ}^*$ minimize
$$\sum_{x,y,z}Q_{XYZ}(x,y,z)\log\frac{Q_{X|YZ}(x|y,z)}{G(x)W(z|x)}$$
over $\calU(Q_{X|Y})$. Observe that
\begin{equation}
\tilde{Q}_{X|YZ}=\left(1-\sqrt{\Delta}-\frac{3\epsilon}{\sqrt{\Delta}}\right)Q_{X|YZ}^*
+\left(\sqrt{\Delta}+\frac{3\epsilon}{\sqrt{\Delta}}\right)Q_X\in\calU(Q_{X|Y},\Delta+3\epsilon)
\end{equation}
since 
\begin{eqnarray}
H_{\tilde{Q}}(X|Y,Z)&\ge&
\left(1-\sqrt{\Delta}-\frac{3\epsilon}{\sqrt{\Delta}}\right)H_{Q^*}(X|Y,Z)+
\left(\sqrt{\Delta}+
\frac{3\epsilon}{\sqrt{\Delta}}\right)H_Q(X)\\
&\ge&\left(\sqrt{\Delta}+\frac{3\epsilon}{\sqrt{\Delta}}\right)\cdot
\sqrt{\Delta}\\
&=&\Delta+3\epsilon,
\end{eqnarray}
and so,
\begin{eqnarray}
& &\min_{Q_{X|YZ}\in\calU(Q_{X|Y},\Delta+3\epsilon)}\sum_{x,y,z}Q_{XYZ}(x,y,z)
\log\frac{Q_{X|YZ}(x|y,z)}{G(x)W(z|x)}\\
&\le&\sum_{y,z}Q_{YZ}(y,z)\sum_x
\tilde{Q}_{X|YZ}(x|y,z)\log\frac{\tilde{Q}_{X|YZ}(x|y,z)}{G(x)W(z|x)}\\
&=&\sum_{y,z}Q_{YZ}(y,z)\sum_x
\tilde{Q}_{X|YZ}(x|y,z)\log\frac{1}{G(x)W(z|x)}-H_{\tilde{Q}}(X|Y,Z)\\
&\le&\left(1-\sqrt{\Delta}-\frac{3\epsilon}{\sqrt{\Delta}}\right)\sum_{y,z}Q_{YZ}(y,z)
\sum_xQ_{X|YZ}^*(x|y,z)\log\frac{1}{G(x)W(z|x)}+\nonumber\\
& &+\left(\sqrt{\Delta}+\frac{3\epsilon}{\sqrt{\Delta}}\right)
\sum_{y,z}Q_{YZ}(y,z)\sum_xQ_X(x)\log\frac{1}{G(x)W(z|x)}-\nonumber\\
& &\left(1-\sqrt{\Delta}-\frac{3\epsilon}{\sqrt{\Delta}}\right)H_{Q^*}(X|Y,Z)-
\left(\sqrt{\Delta}+\frac{3\epsilon}{\sqrt{\Delta}}\right)H_Q(X)\\
&\le&\left(1-\sqrt{\Delta}-\frac{3\epsilon}{\sqrt{\Delta}}\right)
\sum_{y,z}Q_{YZ}(y,z)\sum_x
Q_{X|YZ}^*(x|y,z)\log\frac{Q_{X|YZ}^*(x|y,z)}{G(x)W(z|x)}+\nonumber\\
& &\left(\sqrt{\Delta}+\frac{3\epsilon}{\sqrt{\Delta}}\right)
\sum_{x,z}Q_X(x)Q_Z(z)\log\frac{1}{G(x)W(z|x)}\\
&\le&\left(1-\sqrt{\Delta}-\frac{3\epsilon}{\sqrt{\Delta}}\right)
\min_{Q_{X|YZ}\in\calU(Q_{X|Y})}\sum_{x,y,z}Q_{XYZ}(x,y,z)
\log\frac{Q_{X|YZ}(x|y,z)}{G(x)W(z|x)}+\nonumber\\
& &\left(\sqrt{\Delta}+\frac{3\epsilon}{\sqrt{\Delta}}\right)
\max_{x,z}\log\frac{1}{G(x)W(z|x)}\\
&<&\min_{Q_{X|YZ}\in\calU(Q_{X|Y})}\sum_{x,y,z}Q_{XYZ}(x,y,z)
\log\frac{Q_{X|YZ}(x|y,z)}{G(x)W(z|x)}+\nonumber\\
& &\left(\sqrt{\Delta}+\frac{3\epsilon}{\sqrt{\Delta}}\right)
\max_{x,z}\log\frac{1}{G(x)W(z|x)}.
\end{eqnarray}
Observe that the special case where $X\equiv Y$,
$B_Q(Y,Z)=H_Q(Y,Z)+H_Q(Y|Y,Z)+D(Q_{YZ}\|G\times
W)=H_Q(Y,Z)+D(Q_{YZ}\|G\times W)$, which is suitable also for the case where
$H_Q(Y)<\sqrt{\Delta}$. 
Thus, to summarize, for $f\in\calG$ and $\by\in\calC$, when $\Delta$
(and hence also $\epsilon$) is very small, then essentially, $P(\by)\gexe e^{-nA_Q(Y)}$ and
$P(\by,\bz)\gexe e^{-nB_Q(Y,Z)}$. 
Earlier, we introduced the function $\alpha(\hP_{\by})$ as an
alternative notation that emphasizes the dependence on $\by$. By the same
token, we now introduce the notation $\beta(\hP_{\by\bz})$ and as alternative
to $B_Q(Y,Z)$, for $(\by,\bz)\in\calT(Q_{YZ})$.
Since we have already seen the matching\footnote{Matching -- within
infinitesimally small terms in the exponent.} upper bounds, $P(\by)\lexe e^{-nA_Q(Y)}$ and
$P(\by,\bz)\lexe e^{-nB_Q(Y,Z)}$, in the proof of Theorem 1, then
we observe that for the vast majority of codes
$\{f\}$, the likelihood function (\ref{exactlikelihood}) can be approximated by 
\begin{equation}
P(\bz|\by)\exe \exp\{-n[\beta(\hat{P}_{\by\bz})-\alpha(\hat{P}_{\by})]\}
\dfn e^{-n\gamma(\hat{P}_{\by\bz})},
\end{equation}
whenever $\by\in\calC$. More precisely, in view of the above upper and lower bounds
to $P(\by)$ and $P(\by,\bz)$, we have
\begin{equation}
\label{gammapprox1}
P(\bz|\by)\ge\exp\{-n[\gamma(\hat{P}_{\by\bz})+\theta(\Delta,\epsilon)]-O_1(\log n)-O_2(\log
n)\}
\end{equation}
and
\begin{equation}
\label{gammapprox2}
P(\bz|\by)\le\exp\{-n[\gamma(\hat{P}_{\by\bz})-\Delta-2\epsilon]+O_1(\log n)+O_2(\log
n)\}.
\end{equation}
Thus, a good approximation to the ML decoder, which
achieves the same exponent (in the limit $\epsilon\to 0$ and $\Delta\to 0$) is
given by:
\begin{equation}
\label{approxMLdec}
\hat{m}_{\mbox{\tiny a}}=\mbox{arg min}_m \gamma(\hat{P}_{\by_m\bz}).
\end{equation}
We next derive a lower bound to the average\footnote{Averaging w.r.t.\ the
randomness of $\{\bx_m\}$ while $f\in\calG$ is given.} error probability
of the optimal, ML decoder. 
As in \cite{FL98} and \cite{LaZ98}, to obtain an efficient lower 
bound, we define a tie--breaking mechanism for the ML decoder by means of a ranking
function $M_{\mbox{\tiny o}}(\by,\bz)$, which for a given $\bz$, is a
one--to--one mapping
from $\calC$ to $\{1,2,\ldots,|\calC|\}$, that satisfies the rule
that $P(\bz|\by) > P(\bz|\by')$ implies $M_{\mbox{\tiny o}}(\by,\bz) <
M_{\mbox{\tiny o}}(\by',\bz)$ for every $\by,\by'\in\calC$. Then,
for $f\in\calG$,
\begin{eqnarray}
\mbox{P}_{\mbox{\tiny
e,o}}(f)&\ge&\frac{1}{2}\sum_{\by,\bz}P(\by,\bz)
\min\left\{1,e^{nR_{\mbox{\tiny I}}}\cdot\sum_{\{\by':~M_{\mbox{\tiny
o}}(\by',\bz)\le M_{\mbox{\tiny o}}(\by,\bz)\}\cap\calC}
P(\by')\right\}\nonumber\\
&=&\frac{1}{2}\sum_{\bz}P(\bz)\sum_{\by}P(\by|\bz)
\min\left\{1,e^{nR_{\mbox{\tiny I}}}\cdot\sum_{\{\by':~M_{\mbox{\tiny
o}}(\by',\bz)\le M_{\mbox{\tiny o}}(\by,\bz)\}\cap\calC}
P(\by')\right\}\nonumber\\
&\dfn&\frac{1}{2}\sum_{\bz}P(\bz)\cdot\Pi(\bz),
\end{eqnarray}
where we have used Shulman's
lower bound \cite[Lemma A.2]{Shulman03} 
on the probability of the union of pairwise independent events, relying
on the fact that for a given $f$, the various quantized codewords
$\{\by_m\}$ are independent due to the independence of $\{\bx_m\}$.
Let us also define
\begin{equation}
\tilde{\Pi}(\bz)=\sum_{\by}P(\by|\bz)\cdot\min\left\{1,e^{nR_{\mbox{\tiny I}}}\cdot
\sum_{\{\by':~P(\bz|\by')\ge
e^{-n\delta_n(\Delta,\epsilon)}P(\bz|\by)\}\cap\calC}P(\by')\right\},
\end{equation}
where
$\delta_n(\Delta,\epsilon)=\theta(\Delta,\epsilon)+\Delta+2\epsilon+2[O_1(\log
n)+O_2(\log n)]$.
We show in Subsection A.3 of the appendix (as an extension of \cite[Lemma 2]{LaZ98} and
similarly to \cite[Lemma 1]{p191}) that
\begin{equation}
\label{appendix}
\Pi(\bz)\ge\left[1+e^{n\delta_n(\Delta,\epsilon)}\left(1+n\ln\frac{1}{G_{\min}}\right)\right]^{-1}
\tilde{\Pi}(\bz),~~~~~~\forall \bz\in\calZ^n
\end{equation}
where $G_{\min}\dfn\min_{x\in\calX}G(x)$,
and so, it follows that
\begin{eqnarray}
\mbox{P}_{\mbox{\tiny
e,o}}(f)&\ge&\frac{1}{2}\sum_{\bz}P(\bz)\cdot\Pi(\bz)\\
&\ge&\frac{1}{2}\left[1+e^{n\delta_n(\Delta,\epsilon)}\left(1+n\ln\frac{1}{G_{\min}}\right)\right]^{-1}
\sum_{\bz}P(\bz)\cdot \tilde{\Pi}(\bz)\\
&=&\frac{1}{2}\left[1+e^{n\delta_n(\Delta,\epsilon)}
\left(1+n\ln\frac{1}{G_{\min}}\right)\right]^{-1}
\sum_{\bz}P(\bz)\cdot\sum_{\by\in\calC}P(\by|\bz)\times\nonumber\\
& &\min\left\{1,e^{nR_{\mbox{\tiny I}}}\cdot
\sum_{\{\by':~P(\bz|\by')\ge
e^{-n\delta_n(\Delta,\epsilon)}P(\bz|\by)\}\cap\calC}P(\by')\right\}\\
&\ge&\frac{1}{2}\left[1+e^{n\delta_n(\Delta,\epsilon)}
\left(1+n\ln\frac{1}{G_{\min}}\right)\right]^{-1}
\sum_{\by\in\calC}\sum_{\bz}P(\by,\bz)\times\nonumber\\
& & \min\left\{1,e^{nR_{\mbox{\tiny I}}}\cdot
\sum_{\{\by':~\gamma(\hP_{\by'\bz})\le
\gamma(\hP_{\by\bz})\}\cap\calC}P(\by')\right\}\\
&\ge&\frac{1}{2}\left[1+e^{n\delta_n(\Delta,\epsilon)}
\left(1+n\ln\frac{1}{G_{\min}}\right)\right]^{-1}
\sum_{\by\in\calC}\sum_{\bz}P(\by,\bz)\times\nonumber\\
& &\min\left\{1,e^{nR_{\mbox{\tiny I}}}\cdot
\sum_{\{\by'\in \calT(\by|\bz)\cap\calC\}}P(\by')\right\}\\
&=&\frac{1}{2}\left[1+e^{n\delta_n(\Delta,\epsilon)}
\left(1+n\ln\frac{1}{G_{\min}}\right)\right]^{-1}
\sum_{\by\in\calC}\sum_{\bz}P(\by,\bz)\times\nonumber\\
& &\min\left\{1,e^{nR_{\mbox{\tiny I}}}\cdot
P[\calT(\by|\bz)\cap\calC]\right\}\\
&=&\frac{1}{2}\left[1+e^{n\delta_n(\Delta,\epsilon)}
\left(1+n\ln\frac{1}{G_{\min}}\right)\right]^{-1}
\sum_{\by\in\calC}\sum_{\bz}P(\by,\bz)\times\nonumber\\ 
& &\min\left\{1,e^{nR_{\mbox{\tiny I}}}\cdot
|\calT(\by|\bz)\cap\calC|\cdot P(\by)\right\}\\
&=&\frac{1}{2}\left[1+e^{n\delta_n(\Delta,\epsilon)}
\left(1+n\ln\frac{1}{G_{\min}}\right)\right]^{-1}
\sum_{\by\in\calC}\sum_{\bz}P(\by,\bz)\times\nonumber\\
& & \min\left\{1,e^{nR_{\mbox{\tiny I}}}\cdot
N(\by|\bz)\cdot P(\by)\right\},
\end{eqnarray}
where in the third inequality, we have used the fact that
$\{\by':~\gamma(\hP_{\by'\bz})\le \gamma(\hP_{\by\bz})\}$ is a subset of
$\{\by':~P(\bz|\by')\ge e^{-n\delta_n(\Delta,\epsilon)}P(\bz|\by)\}$, as
implied by eqs.\ (\ref{gammapprox1}) and (\ref{gammapprox2}).
Since the last expression is of the same exponential order as eq.\
(\ref{ubpe}), of the upper bound (after taking $\epsilon$ and $\Delta$ to zero) 
then so is its expectation\footnote{There is a minor issue that has to be kept in
mind when taking the expectation. 
The lower bound for a given $f$ is applicable only for $f\in\calG$, not
for every $f$. But since $\calG^c$ is an extremely small minority of the codes 
(i.e., a double--exponentially small fraction of them), then the
contribution of codes outside $\calG$ can safely be neglected in the
exponential scale, and so, the
expectation over all codes is exponentially the same as the expectation over
all codes within $\calG$.}
w.r.t.\ the randomness of $f$, where
here the above derived (exponentially tight) lower bounds to $P(\by)$ and $P(\by,\bz)$
should be used. This would yield a lower bound to $\bar{\mbox{P}}_{\mbox{\tiny
e,o}}$, which is of the exponential order of $e^{-nE(R_{\mbox{\tiny I}})}$.
This completes the proof of Theorem 2.

\section{Summary and Conclusion}

The main contributions of this work were as follows. We proposed a universal
decoder, which is a variant of the MMI decoder, but is different in the sense
that it takes into account the distribution of the quantized codewords (for a
given lossy source encoder). We analyzed the error exponent of this decoder
and have shown that it improves on the ordinary MMI decoder, analyzed in
\cite{DD11}, and sometimes strictly so. 
We have also shown that our proposed
decoder provides the same error exponent as that of the ML decoder, at least
as long as all single--letter transition probabilities of the channel,
$\{W(z|x)\}$ are strictly positive, and we speculate that this positivity
constraint can be removed. Our decoder is also at least as good as any other
decoder whose decoding metric depends on $(\by_m,\bz)$ only via
the joint empirical distribution $\hP_{\by_m\bz}$.
As a byproduct of our analysis, for a known channel $W$, we have also
proposed a (non--universal) approximate ML decoder (\ref{approxMLdec}), which is
easier to implement than the exact ML decoder, based on
(\ref{exactlikelihood}), yet it yields the same error exponent,
$E(R_{\mbox{\tiny I}})$.

\section*{Appendix}
\renewcommand{\theequation}{A.\arabic{equation}}
    \setcounter{equation}{0}
\subsection*{A.1 Modifying the Map $Q_X\to Q_Y$ To Be One--to--One}

Let $Q_{Y|X}^*=U[Q_X]$ denote our favorite
choice of $Q_{Y|X}$ as a functional of $Q_X$, and let $Q_Y^*=(Q_X\times
Q_{Y|X}^*)_Y\dfn V[Q_X]$. The mapping $V[\cdot]$ may not necessarily
be one--to--one. We would like to perturb $Q_{Y|X}^*$ very slightly
(so that performance would be degraded by a small amount only), 
to $\tQ_{Y|X}$, such that $\tQ_Y=(Q_X\times \tQ_{Y|X})_Y=\tV[Q_X]$
would be one--to--one. We next describe one concrete way to do this.

Without loss of generality, assume the alphabet $\calX$ to be
$\{0,1,\ldots,K-1\}$, where $K=|\calX|$. For convenience, we will also assume that
$|\calY|=|\calX|$, and so, $\calY$ will also be taken to be
$\{0,1,\ldots,K-1\}$ (the extension to the case $|\calY| > |\calX|$
will be straightforward). We first form a fine partition of the simplex. One
way of doing this is the following.
Let $\epsilon > 0$ be arbitrarily small, chosen such that $1/\epsilon$ is integer, and
consider the partition of the simplex $\calQ(\calX)$,
of probability distributions $\{Q_X(x)\}$ over
$\calX$, into cells of size $\epsilon$ such that in each cell,
the letter probabilities are bounded by $i_x\epsilon \le Q_X(x)< (i_x+1)\epsilon$,
$x=1,2,\ldots,K-1$, for some given non--negative integers,
$\{i_x\}_{x=1}^{K-1}$, which will be denoted collectively by $\bi$. Let
$\calQ_{\bi}$ denote the cell pertaining to the index vector $\bi$.
Assuming that $U[\cdot]$ (and hence also $V[\cdot]$) is continuous at least within each
cell (otherwise, form any other fine partition of $\calQ(\calX)$ with this
property), let
$V[\calQ_{\bi}]$ denote the image of $\calQ_{\bi}$ under $V$ and let
$Q_Y^{\bi}$ denote an arbitrary representative member of $V[\calQ_{\bi}]$,
which is taken to have strictly positive letter probabilities (if this is not
the case, then slightly perturb the zero--probabilities to small positive values).
Thus, the number of distinct representatives, $\{Q_{Y|X}^{\bi}\}$, 
cannot exceed the number of cells, which is finite. 
Let $Q_0$ be an arbitrary distribution over $\calY$. 
Now, consider the mapping $\tV$ that maps $Q_X\in\calQ_{\bi}$ to 
$\tQ_Y=Q_Y^{\bi}+\delta\cdot(Q_X-Q_0)$, where $\delta > 0$ is small enough such that
$0\le\tQ_Y(y)\le 1$ for all $y$ and that the sets 
$\{Q_Y^{\bi}+\delta\cdot(Q_X-Q_0):~Q_X\in\calQ(\calX)\}$ are disjoint for
every two
different index vectors $\{\bi\}$ (in particular, $\delta$ should not exceed
$\min_{\bi\ne\bi^\prime}\max_y|Q_Y^{\bi}(y)-Q_Y^{\bi^\prime}(y)|$).
Then, this mapping from $Q_X$ to $\tQ_Y$ is
clearly one--to--one. 
Finally, one can always slightly perturb $Q_{Y|X}^*$ to obtain a new
channel $\tQ_{Y|X}$ such that $(Q_X\times \tQ_{Y|X})_Y=\tQ_Y$, as there are
as many as $K-1$ degrees of freedom to this end.
The perturbations that take us from $Q_Y^*$ to $Q_Y^{\bi}$, and
then to $\tQ_Y$, as well as the perturbation from $Q_{Y|X}^*$ to $\tQ_{Y|X}$, are
arbitrarily small, and hence so is the loss of performance.

\subsection*{A.2 $D(Q_X^*\|G)$ Might Be Strictly Positive}

For simplicity, let us consider the case where $\calX=\calY=\{0,1,\ldots,K-1\}$ 
and there is no compression
constraint, so $Q_{Y|X}$
can be taken to be the identity matrix 
(clearly, this situation can be approached in our setting, in the limit where the compression
constraints are sufficiently soft) and let $R_{\mbox{\tiny I}}=0$. 
Suppose further that $W$ is also the identity matrix, i.e., the clean channel
(which again, can be thought of as a limit of very good channels).
In this case, $E(0)$ simplifies to
\begin{equation}
E(0)=\min_{Q_X}[2D(Q_X\|G)+H_Q(X)],
\end{equation}
which is easily shown to be achieved by
\begin{equation}
Q_X^*(x)=\frac{G^2(x)}{\sum_{x^\prime\in\calX}G^2(x^\prime)},
\end{equation}
that differs from $G$ (except some special cases)
and hence $D(Q_X^*\|G)>0$. On substituting $Q_X^*$ back into the
expression of $E(0)$, we obtain 
\begin{equation}
E(0)=-\log\left[\sum_xG^2(x)\right],
\end{equation}
as expected. On
the other hand, the error exponent of \cite{DD11}, in this case, becomes
\begin{equation}
E_{\mbox{\tiny DD}}(0)=
\min_{Q_X}[D(Q_X\|G)+H_Q(X)]=\min_{Q_X}\sum_xQ(x)\log\frac{1}{G(x)}=-\log
\left[\max_xG(x)\right],
\end{equation}
which is always smaller, except for some special cases.
The same gap continues to apply at least for a certain range of low rates, where
$E(R_{\mbox{\tiny I}})= E(0)-R_{\mbox{\tiny I}}$ 
and $E_{\mbox{\tiny DD}}(R_{\mbox{\tiny I}})=E_{\mbox{\tiny
DD}}(0)-R_{\mbox{\tiny I}}$.

\subsection*{A.3 Proof of Eq.\ (\ref{appendix})}

The proof is very similar to the proof of Lemma 1 in \cite{p191}, which in
turn, is an extension of \cite[Lemma 2]{LaZ98}, and it is given here for the
sake of completeness.
For brevity, let us denote $\alpha=e^{n\delta_n(\Delta,\epsilon)}$
and define
\begin{eqnarray}
\Delta(\by,\bz)
&\dfn&\{\by':~M_{\mbox{\tiny o}}(\by',\bz)> M_{\mbox{\tiny o}}(\by,\bz),
~P(\bz|\by')\ge \alpha^{-1}P(\bz|\by)\}\cap\calC\\
&=&\{\by':~M_{\mbox{\tiny o}}(\by',\bz)> M_{\mbox{\tiny o}}(\by,\bz),
~P(\by)P(\by'|\bz)\ge \alpha^{-1}P(\by')P(\by|\bz)\}\cap\calC,
\end{eqnarray}
so that 
$$\calE_{\mbox{\tiny t}}(\by,\bz)\dfn
\{\by':~P(\bz|\by')\ge \alpha^{-1}P(\bz|\by)\}\cap\calC$$
is given by the disjoint union of $\Delta(\by,\bz)$ and
$$\calE_{\mbox{\tiny o}}(\by,\bz)\dfn\{\by':~M_{\mbox{\tiny o}}(\by',\bz)<
M_{\mbox{\tiny o}}(\by,\bz)\}\cap\calC.$$ 
Let us also define the function $\phi(t)=\min\{1,t\cdot e^{nR_{\mbox{\tiny
I}}}\}$ for $t\ge 0$, and
observe that for $t\le s$, $\phi(s)\le \frac{s}{t}\cdot\phi(t)$, as can easily
be seen
from the concavity of $\phi(\cdot)$ and the fact that $\phi(0)=0$.
Thus,
\begin{eqnarray}
\Pi(\bz)&=&\sum_{\by}P(\by|\bz)\phi(P[\calE_{\mbox{\tiny
o}}(\by,\bz)])\\
\tilde{\Pi}(\bz)&=&\sum_{\by}P(\by|\bz)\phi(P[\calE_{\mbox{\tiny
o}}(\by,\bz)]+P[\Delta(\by,\bz)])\\
&\le&\sum_{\by}P(\by|\bz)\left(\frac{P[\calE_{\mbox{\tiny o}}(\by,\bz)]+
P[\Delta(\by,\bz)]}{(P[\calE_{\mbox{\tiny
o}}(\by,\bz)]}\right)\phi(P[\calE_{\mbox{\tiny o}}(\by,\bz)]),
\end{eqnarray}
where in the last inequality, we have used the above mentioned property of
the function $\phi(\cdot)$.
Now, let us define
\begin{equation}
r(\by,\bz)\dfn\sum_{\by'\in\calE_{\mbox{\tiny o}}(\by,\bz)}P(\by'|\bz).
\end{equation}
Then, for $\by\in\calC$,
\begin{eqnarray}
P(\by)&=&\sum_{\by'}P(\by)P(\by'|\bz)\\
&\ge&\sum_{\by'\in\calE_{\mbox{\tiny o}}(\by,\bz)}P(\by)P(\by'|\bz)+
\sum_{\by'\in\Delta(\by,\bz)}P(\by)P(\by'|\bz)\\
&=&P(\by)r(\by,\bz)+
\sum_{\by'\in\Delta(\by,\bz)}P(\by)P(\by'|\bz)\\
&\ge&P(\by)r(\by,\bz)+\frac{1}{\alpha}
\sum_{\by'\in\Delta(\by,\bz)}P(\by')P(\by|\bz)\\
&=&P(\by)r(\by,\bz)+\frac{P(\by|\bz)}{\alpha}
P[\Delta(\by,\bz)],
\end{eqnarray}
and so,
\begin{equation}
P(\by|\bz)P[\Delta(\by,\bz)]\le \alpha P(\by)[1-r(\by,\bz)].
\end{equation}
We then have
\begin{eqnarray}
& &\tilde{\Pi}(\bz)-\Pi(\bz)\\
&\le&\sum_{\by\in\calC}P(\by|\bz)\frac{P[\Delta(\by,\bz)]}
{P[\calE_{\mbox{\tiny o}}(\by,\bz)]}\phi(P[\calE_{\mbox{\tiny o}}(\by,\bz)])\\
&\le&\alpha\cdot\sum_{\by\in\calC}\frac{P(\by)[1-r(\by,\bz)]}
{P[\calE_{\mbox{\tiny o}}(\by,\bz)]}\phi(P[\calE_{\mbox{\tiny o}}(\by,\bz)])\\
&=&\alpha\cdot\sum_{\by\in\calC}\sum_{\{\by'\in\calC:~M_{\mbox{\tiny o}}(\by',\bz) >
M_{\mbox{\tiny o}}(\by,\bz)\}}\frac{P(\by)P(\by'|\bz)}
{P[\calE_{\mbox{\tiny o}}(\by,\bz)]}\phi(P[\calE_{\mbox{\tiny o}}(\by,\bz)])\\
&\eqa&\alpha\cdot\sum_{\by'\in\calC}\sum_{\{\by\in\calC:~M_{\mbox{\tiny o}}(\by',\bz) >
M_{\mbox{\tiny
o}}(\by,\bz)\}}
\frac{P(\by)P(\by'|\bz)}{P[\calE_{\mbox{\tiny
o}}(\by,\bz)]}\phi(P[\calE_{\mbox{\tiny o}}(\by,\bz)])\\
&\leb&\alpha\cdot\sum_{\by'\in\calC}\sum_{\{\by\in\calC:~M_{\mbox{\tiny o}}(\by',\bz) >
M_{\mbox{\tiny
o}}(\by,\bz)\}}
\frac{P(\by)P(\by'|\bz)}{P[\calE_{\mbox{\tiny o}}(\by,\bz)]}
\phi(P[\calE_{\mbox{\tiny o}}(\by',\bz)])\\
&\le&\alpha\cdot\sum_{\by'\in\calC}P(\by'|\bz)\phi(P[\calE_{\mbox{\tiny
o}}(\by',\bz)])\cdot\sum_{\by}
\frac{P(\by)}{P[\calE_{\mbox{\tiny o}}(\by,\bz)]}\\
&=&\alpha\cdot\Pi(\bz)
\cdot\sum_{\by\in\calC}\frac{P(\by)}{P[\calE_{\mbox{\tiny
o}}(\by,\bz)]},
\end{eqnarray}
where in (a) we have interchanged the order of the summation and in (b), we
have used the monotonicity of $\phi(\cdot)$ together with the fact that
$\calE_{\mbox{\tiny o}}(\by,\bz)\subseteq \calE_{\mbox{\tiny o}}(\by',\bz)$
whenever $M_{\mbox{\tiny o}}(\by',\bz) > M_{\mbox{\tiny o}}(\by,\bz)$.
To complete the proof, it remains to show then that
for any $\bz$,
\begin{equation}
K_n(\bz)\dfn
\sum_{\by\in\calC}\frac{P(\by)}{P[\calE_{\mbox{\tiny o}}(\by,\bz)]}=
\sum_{\by\in\calC}\frac{P(\by)}{\sum_{\{\by':~M_{\mbox{\tiny o}}(\by',\bz)
\le
M_{\mbox{\tiny o}}(\by,\bz)\}}P(\by')}
\end{equation}
cannot exceed $1+n\ln(1/G_{\min})$.
For the given $\bz$, consider the ordering of all members of $\calC$
according to the ranking function $M_{\mbox{\tiny o}}(\by,\bz)$, i.e.,
\begin{equation}
P(\bz|\by[1])\ge
P(\bz|\by[2])\ge\ldots\ge P(\bz|\by[N]),~~~~~~~~N=|\calC|
\end{equation}
and let us denote $a_i=P(\by[i])$, $A_i=\sum_{j=1}^i a_j$, $i=1,\ldots,N$.
Then, using the facts that $A_1=a_1=P(\by[1])$ and $A_N=1$, as well as the
inequality
\begin{equation}
\ln(1+u)\equiv-\ln\left(1-\frac{u}{1+u}\right)\ge\frac{u}{1+u},
\end{equation}
we have
\begin{eqnarray}
K_n(\bz)&=&\sum_{i=1}^N\frac{a_i}{A_i}\\
&=&1+\sum_{i=2}^N\frac{a_i}{A_{i-1}+a_i}\\
&=&1+\sum_{i=2}^N\frac{a_i/A_{i-1}}{1+a_i/A_{i-1}}\\
&\le&1+\sum_{i=2}^N\ln\left(1+\frac{a_i}{A_{i-1}}\right)\\
&=&1+\sum_{i=2}^N\ln\left(\frac{A_{i-1}+a_i}{A_{i-1}}\right)\\
&=&1+\sum_{i=2}^N\ln\left(\frac{A_i}{A_{i-1}}\right)\\
&=&1+\ln\left(\frac{A_N}{A_1}\right)\\
&=&\ln\left[\frac{1}{P(\by[1])}\right]+1\\
&\le&\ln\left(\frac{1}{G_{\min}^n}\right)+1\\
&=&n\ln\left(\frac{1}{G_{\min}}\right)+1,
\end{eqnarray}
where we have used the fact that for every code in $\calG$, 
and $\by\in\calC$, $P(\by)> 0$, and it is at least as large as $G(\bx)$
for some $\bx\in f^{-1}(\by)$, which in turn, cannot be less than
$G_{\min}^n$.
This completes the proof of eq.\ (\ref{appendix}).


\begin{thebibliography}{AA}

\bibitem{CK11}
I.~Csisz\'ar and J.~K\"orner, {\it Information Theory: Coding Theorems for
Discrete Memoryless Systems}, Cambridge University Press, 2011.

\bibitem{DD11}
G.~Dasarathy and S.~C.~Draper, ''On reliability of content identification from
databases based on noisy queries,''
{\it The 2011 IEEE Proc.\ International Symposium on
Information Theory (ISIT 2011)}, pp.\ 1066--1070, St.\ Petersburg, Russia,
July--August 2011.

\bibitem{DD14}
G.~Dasarathy and S.~C.~Draper, ''Upper and lower bounds on the reliability of
content identification,'' {\it Proc.\ International Zurich Seminar (IZS)},
pp.\ 100--103, February 2014.

\bibitem{FL98}
M.~Feder and A.~Lapidoth, ''Universal decoding for channels with memory,''
{\em IEEE Trans.~Inform.~Theory\/},
vol.~44, no.~5, pp.~1726--1745, September 1998.

\bibitem{IW10}
T.~Ignatenko and F.~M.~J.~Willems, ''Biometric security from an
information--theoretical perspective,'' {\it Foundations and Trends in
Communications and Information Theory}, vol.\ 7, nos.\ 2--3, pp.\ 135--316.

\bibitem{LaZ98}
A.~Lapidoth and J.~Ziv, ''On the universality of the LZ--based noisy
channels decoding algorithm,''
{\em IEEE Trans.~Inform.~Theory\/},
vol.~44, no.~5, pp.~1746--1755, September 1998.

\bibitem{fnt}
N.~Merhav, ``Statistical physics and information theory,''
(invited paper) {\it Foundations and Trends in
Communications and Information Theory}, vol.\ 6, nos.\ 1--2, pp.\ 1--212,
2009.

\bibitem{givenclassofmetrics}
N.~Merhav, ''Universal decoding for arbitrary channels
relative to a given family of decoding metrics,''
{\it IEEE Trans.\ Inform.\ Theory}, vol.\ 59, no.\ 9, pp.\ 5566--5576,
September 2013.

\bibitem{p191}
N.~Merhav, ''Universal decoding using a noisy codebook,'' submitted for
publication and available on--line at:
{\tt http://arxiv.org/pdf/1609.00549.pdf}

\bibitem{Shulman03}
N.~Shulman, {\it Communication over an Unknown Channel via
Common Broadcasting}, Ph.D.\ dissertation, Department of Electrical
Engineering -- Systems, Tel Aviv University, July 2003.\\
http://www.eng.tau.ac.il/$\sim$shulman/papers/Nadav\_PhD.pdf

\bibitem{Tuncel09}
E.~Tuncel, ''Capacity/storage tradeoff in high--dimensional identification
systems,'' {\em IEEE Trans.~Inform.~Theory\/}, vol.~55, no.~5, pp.~2097--2106,
May 2009.

\bibitem{VW09}
A.~L.~Varna and M.~Wu,
''Modeling and analysis of content identification,''
{\it Proc.\ 2009 IEEE International Conference on Multimedia and Expo (ICME
2009)}, pp.\ 1528--1531, New York, U.S.A., June--July 2009.

\bibitem{WO08}
M.~B.~Westover and J.~A.~O'Sullivan, ''Achievable rates for pattern
recognition,'' {\em IEEE Trans.~Inform.~Theory\/}, vol.~54, no.\ 1, pp.\ 299--320,
January 2008.

\bibitem{WKGL03}
F.~Willems, T.~Kalker, J.~Goseling, and J.-P.~Linnartz, ''On the capacity of a
biometrical identification system,'' {\it The 2003 IEEE Proc.\ International
Symposium on
Information Theory (ISIT 2003)}, p.~82, Yokohama, Japan, June--July 2003.

\bibitem{WKBL}
F.~Willems, T.~Kalker, S.~Baggen, and J.-P.~Linnartz, ''On the capacity of a
biometrical identification system,'' (unknown year) available on--line at:\\
{\tt http://citeseerx.ist.psu.edu/viewdoc/download?doi=10.1.1.74.9512\&rep=rep1\&type=pdf}

\end{thebibliography}
\end{document}